\documentclass[preprint,5p,times,twocolumn]{elsarticle}

\usepackage[english]{babel}
\usepackage{graphicx}
\usepackage{amssymb}
\usepackage{amsmath}
\usepackage{amsfonts}
\usepackage{gensymb}
\usepackage{color}
\usepackage{soul}
\usepackage[update,prepend]{epstopdf}
\usepackage{multicol}

\journal{Soft Matter}

\begin{document}

\begin{frontmatter}

\title{Tuning the order of colloidal monolayers: assembly of heterogeneously charged colloids close to a patterned substrate}

\author[label1]{Emanuele Locatelli}
\author[label1]{Emanuela Bianchi}
\ead{emanuela.bianchi@univie.ac.at}
\address[label1]{Faculty of Physics, University of Vienna, Boltzmanngasse 5, A-1090, Vienna, Austria}

\begin{abstract}
We study the behavior of negatively charged colloids with two positively charged polar caps close to a planar patterned surface. The competition between the different anisotropic components of the particle-particle interaction patterns is able by itself to give rise to a rich assembly scenario: colloids with charged surface patterns form different crystalline domains when adsorbed to a homogeneously charged substrate. Here we consider substrates composed of alternating (negative/neutral, positive/neutral and positive/negative) parallel stripes and, by means of Monte Carlo simulations, we investigate the ordering of the colloids on changing the number of the stripes. We show that the additional competition between the two different lengths scales characterizing the system ($i.e.,$ the particle interaction range and the size of the stripes) gives rise to a plethora of distinct particle arrangements, where some well-defined trends can be observed. By accurately tuning the substrate charged motif it is possible to, $e. g.,$ promote specific particles arrangements, disfavor crystalline domains or induce the formation of extended, open clusters.
\end{abstract}

\end{frontmatter}

\section{Introduction}\label{sec:introduction}

In the vast realm of nano- and micro-scale materials, low dimensional systems, such as colloidal mono- and bi-layers, are relevant for a broad range of applications as, $e.g.$, for antireflection coatings, biosensors, data-storage, optical and photovoltaic devices, or catalysts~\cite{NanoparticleArrays2000,BiosensorVelev,Lithography2012,Coatings2012}. The ordering of the colloidal particles on a surface strongly depends on a delicate balance between the properties of the particles and those of the substrate.  An efficient method for the fabrication of two-dimensional colloidal structures consists, $e.g.$, in tuning the surface properties of the colloidal particles~\cite{Kagomeexp_2011,Pine_2012}. The guided growth of planar architectures can be also achieved by taking advantage of chemically or physically patterned substrates~\cite{review-substrates,stripe-substrates1,stripe-substrates2,stripe-substrates3,bechinger2007}. Here we combine these two approaches by considering both particles and substrates with a charged surface pattern, $i.e.$, we investigate the behavior of non-homogeneously charged colloids  close to a non-homogeneously charged substrate.

Heterogeneously charged colloids have recently emerged as promising self-assembling units for material architectures with target properties~\cite{Yigit15a,Yigit15b,vanOostrum2015,Yura2015-2,Cruz_2016,Blanco,Hiero_2016,Lund2017,Mani2017}. Particles with distinct charged regions on their surface can be generally regarded as charged patchy units;  in order to distinguish them from conventional patchy particles~\cite{patchyrevexp,patchyrevtheo,newreview}, they are often referred to as inverse patchy colloids (IPCs)~\cite{bianchi:2011}. Similar to conventional patchy colloids, IPCs are characterized by anisotropic interaction patterns and low bonding valence, but the competition between the orientation-dependent attraction and repulsion -- induced by the interactions between like/oppositely charged areas on the particle surface -- leads to more complex assembly scenarios as compared to conventional patchy units~\cite{ipc-review,newreview}. A review about this class of systems, covering the synthesis of model particles, their self-assembly, their coarse-grained modeling and the related numerical/analytical treatments, has been presented in Ref.~\cite{ipc-review}. 

Within this broad class of systems, most of the attention has been devoted so far to IPCs with two charged polar patches and an oppositely charged equatorial belt. The characteristic feature of the bonding between IPCs with two identical patches is the direct contact between the polar region of one particle with the equatorial area of the other. This preferred pattern favors the formation of planar aggregates either as monolayers close to a homogeneously charged substrate~\cite{bianchi:2d2013,bianchi:2d2014} or as bulk equilibrium phases~\cite{ismene,evaemanJPCM,silvanonanoscale}.

The assembly of heterogeneously charged colloids close to a homogeneously charged substrates depends on different parameters, such as the particle/substrate charge ratio, the extension of the charged regions on the particle surface, and the net particle charge~\cite{bianchi:2d2013,bianchi:2d2014}. The resulting surface layers can have different densities: sometimes particles assemble into close-packed, hexagonally ordered crystalline aggregates, sometimes they form open, square-like layers, sometimes they form a monomer-phase where particles are adsorbed on the substrate~\cite{bianchi:2d2013,bianchi:2d2014}.  Moreover, upon subtle changes of external parameters, such as the pH of the solution or the electrical charge of the substrate, it is possible to reversibly switch the assembly process on and off as well as to induce a transformation from a one specific particle arrangement to another~\cite{bianchi:2d2014}. The same morphological features observed in simulations are found in experimental samples of IPCs sedimented on a glass substrate~\cite{vanOostrum2015}.  A qualitative comparison between simulations and experiments has been reported in Ref.~\cite{ipc-review}: while in numerical samples all clusters have the same spatial and orientational order, in the experimental sample different particle arrangements coexist, probably because of the patch size polydispersity of the IPCs synthesized so far.

Here we investigate how the ordering of IPCs on a charged surface is affected by the substrate pattern. In particular, we consider negatively charged colloids with positively charged polar caps and substrates characterized by  parallel stripes forming an alternating charge pattern: negative/neutral, positive/neutral and positive/negative. On varying the size of the stripes, we observe the emergence of different particle ordering, from adsorbed monomers to adsorbed crystalline layers with distinct particle arrangements. We also investigate the robustness of the emerging scenarios with respect to changes in the substrate surface charge and in the electrostatic screening conditions. 

\section{Model and Methods}\label{sec:model&methods}

We consider IPCs with two positively charged polar caps and a negatively charged equatorial belt. The polar patches are identical, $i.e.$, they have the same size as well as the same charge. We mostly focus on neutral IPCs, $i.e.$, heterogeneously charged particles with a zero net charge; we also consider charged IPCs carrying a negative net charge. Particles are placed under confinement between two parallel walls at a distance such that two particles cannot sit on top of each other; the top wall is always neutral, while the bottom wall is patterned with parallel stripes, which can be either neutral or positively/negatively charged.

We note that the Debye-H\"uckel approximation behind our coarse-grained description is strictly valid when $ Z_i q_e \Phi \ll k_BT$, where $Z_i q_e$ is the charge of the ionic species (where $Z_i=1$ for monovalent ions and $q_e$ is the elementary charge), $k_B T$ at room temperature is $\approx 26$ meV, and $\Phi$ is the potential at the surface. For IPC systems, by varying the different parameters that characterize the potential (namely, the colloidal size, the effective charges of the different surface areas, the geometric parameters defining the charge distribution and the Debye screening length), it is possible to determine in which parameter windows the linear approximation is expected to be reliable.  In our specific case, $q_e \beta \Phi \lesssim 1$ for $\sigma \ge 30$ nm.  For particles with $\sigma \approx 30$ nm, the salt concentration corresponding to our $\kappa\sigma$ is $\approx 10$ mM.

\begin{figure}[htbp]
	\begin{center}
	\includegraphics[width=0.5\textwidth]{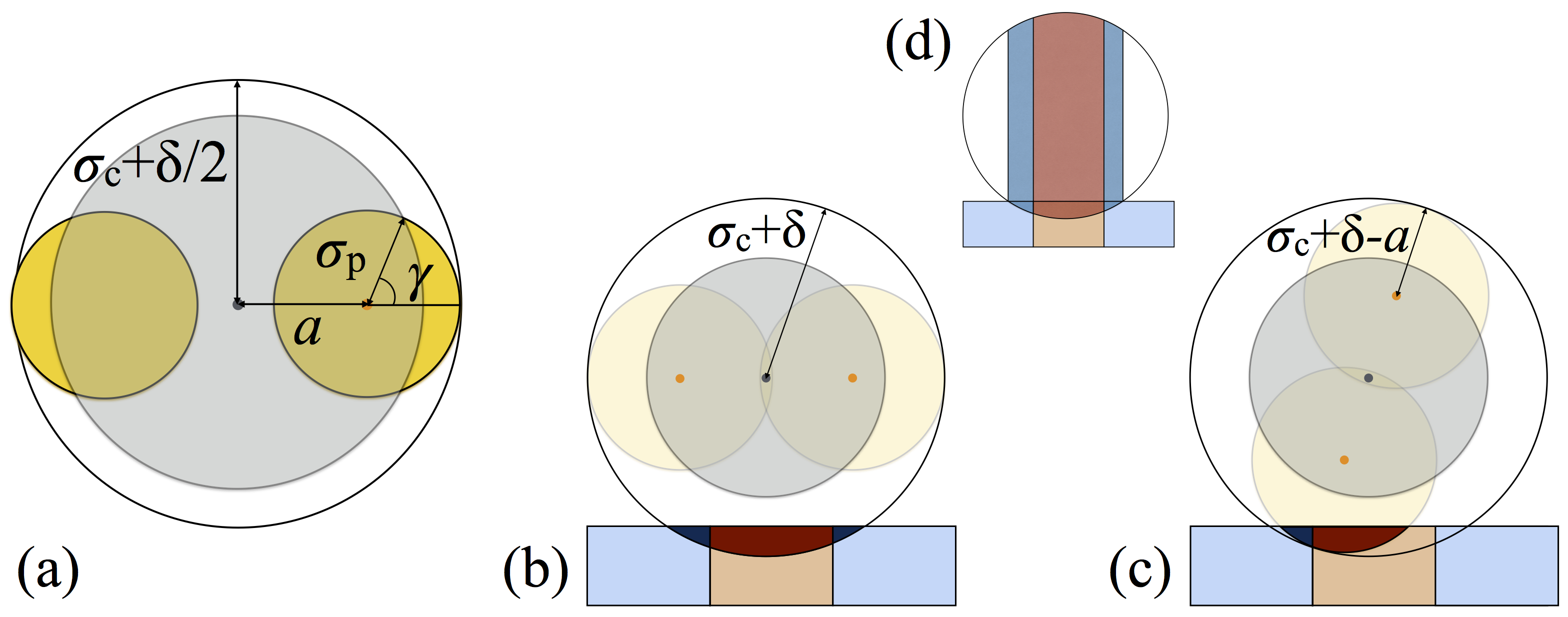} 
	\end{center}
	\caption{Panel (a): schematic representation of a coarse-grained IPC. The gray sphere corresponds to the hard colloid, while the yellow caps represent the polar patches. It is worth noting that IPCs are spherical particles: the yellow caps are the emerging parts of the patch interaction spheres, whose centers are located inside the colloid. The model geometric parameters are the radius of the colloid $\sigma_{\rm c}$, the particle interaction range $\delta$ and the patch opening angle $\gamma$ (as labeled); the latter two are determined once the position of the patch centers $a$ and the patch interaction range $\sigma_{\rm p}$ (as labeled) are chosen. Panels (b) and (c): schematic representation of the interaction between an IPC and a patterned substrate. The different colors correspond to differently charged stripes on the substrate. The intersection between the (core or patch) interaction sphere and the substrate is a spherical cap: the sub-volumes of the caps are highlighted for both the core-substrate (b) and the patch-substrate (c) interaction. Panel (d): particle slices corresponding to the different sub-volumes highlighted in panel (b).}
	\label{fig:model}
\end{figure}

\subsection{Particle-particle interaction}
\label{sec:ppint}
For the particle-particle interaction, we consider the coarse-grained description introduced in Ref.~\cite{bianchi:2011}.  The model features a spherical, impenetrable colloid, of radius $\sigma_{\rm c}$ and central charge $Z_{\rm c}$, carrying two polar patches, each of charge $Z_{\rm p}$. The particle diameter $\sigma=2\sigma_{\rm c}$ sets the unit of length. The model has three sets of independent parameters: the particle interaction range $\delta$, the patch surface extent, specified by the half opening angle $\gamma$, and the interaction strengths between the different surface regions (see panel (a) Fig.~\ref{fig:model}). The pair potential is a sum of three different contributions: the patch/patch, the patch/core and the core/core interactions. The specific form of the potential is based on the postulate that all the interactions can be factorized into a characteristic energy strength and a geometric weight factor, the latter one taking into account the distance and relative orientation between the interacting particles. More specifically, beyond the hard core repulsion, the pair potential between two IPCs at distance $r_{ij}$ with a mutual orientation $\Omega_{ij}$ is given by
\begin{equation}\label{eq:U}
U (r_{ij},\Omega_{ij})  = 
\left\{
\begin{array}{rl}
\sum_{\alpha\beta} u_{\alpha\beta} w_{\alpha\beta}(r_{ij},\Omega_{ij}) &{\hspace{1em}\rm if\hspace{1em}} \sigma <r_{ij}<\sigma+\delta \\
0                          &{\hspace{1em}\rm if\hspace{1em}} r_{ij} \ge \sigma+\delta
\end{array}
\right. 
\end{equation}
where $i$ and $j$ specify the particles, $\alpha$ and $\beta$ specify either a patch (P) or the core (C) of the first and second IPC, respectively, $w_{\alpha\beta}$ and $u_{\alpha\beta}$ are the geometric weight factor and the characteristic energy strength of the $\alpha\beta$ interaction, respectively. We note that, while the $u_{\alpha\beta}$ are constant values, the $w_{\alpha\beta}$ depend on both the inter-particle distance and the relative orientation of the two IPCs. For sake of simplicity we omit in the following the explicit dependence of the $w_{\alpha\beta}$  on $r_{ij}$ and $\Omega_{ij}$. The weight functions for a pair of particles in a given configuration are calculated as overlap volumes between pairs of (core and/or patch) interaction spheres~\cite{bianchi:2011}. The energy strengths are defined $via$ a mapping between a suitably developed Debye-H{\"uckel} potential~\cite{bianchi:2011,debyehueckel} -- fully determined by the physical properties of the underlying microscopic system -- and the patchy pair potential -- characterized by the yet undetermined $u_{\alpha\beta}$~\cite{bianchi:2011}.

Here we consider two IPC types: they share the same patch size and interaction range, while having different net particle charge. As the selected systems have been studied before~\cite{bianchi:2d2014}, we use the names already given in Ref.~\cite{bianchi:2d2014}  for the sake of consistency. In the contest of Ref.~\cite{bianchi:2d2014} it was important to distinguish between IPCs with different patch size and different overall particle charge, thus the systems were labelled as 60n and 60c, where 60 referred to $\gamma$, while ``n" and ``c" indicated neutral and charged particles, respectively. In the following we give all the parameters that characterize both systems.

The interaction range is fixed by the electrostatic screening conditions of the surrounding solvent. We choose $\kappa\sigma=10$ and we define $\delta$ as a function of the Debye screening length $\kappa^{-1}$. While usually $\kappa\delta=1$, we extend the relation between $\delta$ and $\kappa$ to allow for a more quantitative evaluation of the characteristic interaction distance~\cite{bianchi:2011}, $i. e.,$  we assume $\kappa\delta=n$, where $n$ is not necessarily an integer number. Here we choose $n=2$ corresponding to $\delta=0.2$ in units of the particle diameter. 

The patch size is ideally determined by the corresponding feature of experimentally synthesized particles. Once $\delta$ is chosen, $\gamma$ is fixed either by the position of the patch center of charge, $a$, or by the size of the patch interaction sphere $\sigma_{\rm p}$, since the following relations must be satisfied (see panel (a) of Figure~\ref{fig:model}):
\begin{eqnarray}
\frac{\delta}{2}&=&a+\sigma_{\rm p}-\sigma_{\rm c}  \nonumber \\
\cos\gamma&=&\frac{\sigma_{\rm c}^2+a^2-\sigma_{\rm p}^2}{2\sigma_{\rm c} a}.  
\end{eqnarray}
\label{eq:params}
We choose $\gamma \approx 60 \degree$, corresponding to $a = 0.16$, with $\sigma_{\rm p}= 0.44$, both in units of the particle diameter. 

Finally, the energy parameters of the model, $i.e.$, the $u_{\alpha\beta}$ in equation (\ref{eq:U}), are related to the effective charges of the different surface regions. These charges are responsible for the ratio between the attractive and repulsive contributions to the pair energy.  We mostly focus on neutral IPCs, $i.e.$, particles with a net charge $Z_{\rm tot}=Z_{\rm c}+2Z_{\rm p} = 0$. Specifically, we set $Z_{\rm c}=-180$ and $Z_{\rm p}=90$, where we have set the elementary charge to one. We also consider overall charged colloids, such that the net particle charge is negative, namely $Z_{\rm tot}=-\frac{2}{9} Z_{\rm p}<0$ with $Z_{\rm c}=-200$ and $Z_{\rm p}=90$.  To derive the energy strengths  $u_{\alpha\beta}$ in equation (\ref{eq:U}) we perform the ``max" mapping between the model potential of equation (\ref{eq:U}) and the Yukawa-like pair potential appropriately derived within the Debye-H{\"u}ckel approach for IPCs in water at room temperature~\cite{bianchi:2011}. Namely, we match the value of the two potentials for a pair of IPCs at contact ($i.e.$, the distance between the two cores is $r = \sigma$)  with a fixed mutual orientation (shown in Fig.~\ref{fig:configs}). For each $\alpha\beta$ interaction we choose a reference particle configuration~\cite{bianchi:2011}. For the 60n system, we obtain $u_{\rm CC}= 0.1349$ for the core/core interaction, $u_{\rm PC}= -0.8483$  for the patch/core interaction, and $u_{\rm PP}= 4.3228$  for the patch/patch interaction. For the 60c system, we obtain $u_{\rm CC}=  0.4330$, $u_{\rm PC}= -1.9467$, and $u_{\rm PP}= 4.3228$. The pair interaction energy is normalized by the value corresponding to the minimum of the attraction, $\varepsilon_{\rm min}$; thus $|\varepsilon_{\rm min}|$ sets the energy unit. 

The characteristic pair configurations  -- reported in Fig.~\ref{fig:configs} -- are referred to as equatorial-polar (EP), equatorial-equatorial (EE) and polar-polar (PP). 
 We observe that, by construction, the equatorial-polar attraction for two particles at contact is alway -1 in both systems. In contrast, the strengths of the equatorial-equatorial  and polar-polar repulsive interactions for two particles at contact are different in the two systems. For system 60n the polar-polar repulsion is the strongest interaction, while the equatorial-equatorial repulsion is the weakest; for system 60c  the attraction is stronger than both repulsions, which in turn have comparable strength.  A representation of the inter-particle potentials is reported in the Supporting Information of Ref.~\cite{bianchi:2d2014}, while contact energy values for the particle-particle configurations reported in Fig.~\ref{fig:configs} are listed in Tab.~\ref{tab:contactenergies}.

\subsection{Particle-substrate interaction}

\begin{figure}[htbp]
	\begin{center}
	\includegraphics[width=0.4\textwidth]{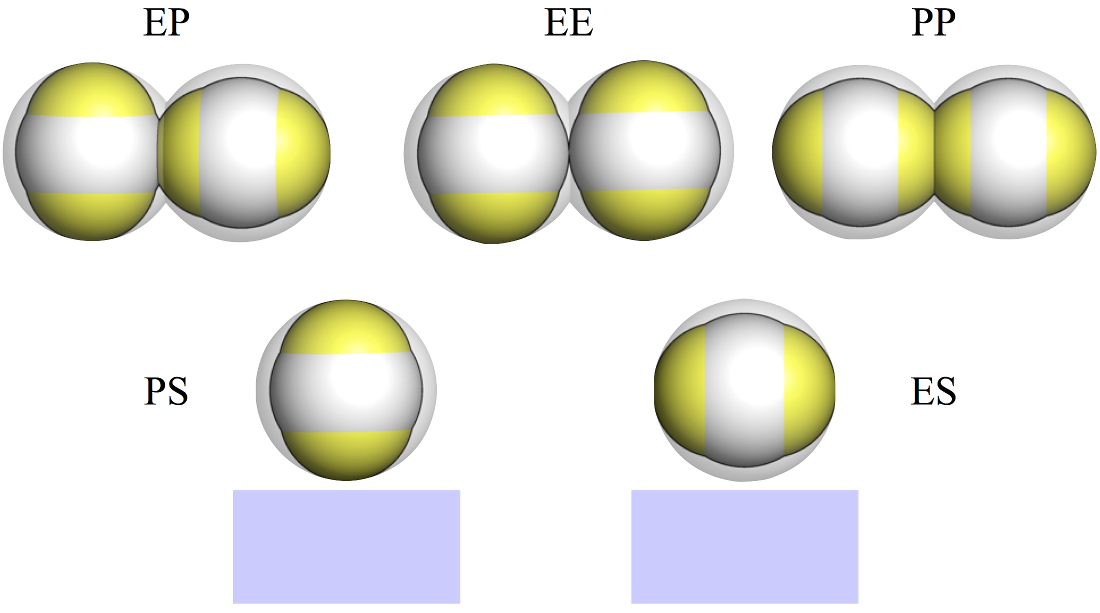} 
	\end{center}
	\caption{Typical particle-particle (top row) and particle-wall (bottom row) configurations: (from left to right, from top to bottom) equatorial-polar (EP), equatorial-equatorial (EE), polar-polar (PP), equatorial-substrate (ES) and patch-substrate (PS), as labeled.}
	\label{fig:configs}
\end{figure}

We consider a system of IPCs confined between two hard walls in a quasi-2D slab along the $xy$-plane, the two walls are positioned at $z =$ 0 and $z =$ 1.45$\sigma$. Such a tight confinement prevents particles from sitting on top of each other. The top wall is always neutral, while the bottom wall -- the substrate -- has a charged pattern: we consider stripes of the same width along the $x$ direction, spanning the plane along the $y$ axis. The stripes form alternating charged motifs: we alternate negative and neutral (-/0), positive and neutral (+/0) or positive and negative (+/0) stripes. We consider different numbers of stripes, $N_s$, spanning from $N_s =$ 2 to $N_s =$ 90, considering only even numbers of stripes, so that the total area covered by each of the two kinds is the same for any choice of $N_s$. We note that, since the stripes fill the whole surface, a low number of stripes implies that each stripe is quite large, the width of a stripe being defined as $L_s = L/N_s$. As $N_s$ increases, stripes become thinner and thinner, until their size becomes smaller than the particle size. At that point, each particle can interact with multiple stripes at the same time: for our choice of the system size, this scenario occurs at $N_s =$ 50. 
For the particle-substrate interaction, we consider the coarse-grained description introduced in Ref.~\cite{bianchi:2d2013}. The interaction between a coarse-grained IPC and a neutral substrate is a steric interaction modeled as a hard repulsion taking place when the particle is located at distance ${\rm z}\le\sigma_{\rm c}$ from the substrate. In the presence of a charged substrate with surface charge $\sigma_{\rm w}=Z_{\rm w}/4\sigma_{\rm c}^2$, a screened electrostatic interaction must be added to the hard particle-substrate repulsion. We consider either positive and negative substrates with a very small value of $Z_{\rm w}$ as compared to $Z_{\rm p}$, $i.e.$, $Z_{\rm w}=\pm\frac{5}{90}Z_{\rm p}=\pm$5. 
Following Ref.~\cite{bianchi:2d2013}, the particle-substrate interaction is modeled according to the same coarse-grained philosophy introduced in Sec.~\ref{sec:ppint}: beyond the hard core repulsion, the interaction energy of an IPC at distance $z_{i}$ and orientation $\Omega_{i}$ with respect to the substrate (S) is given by
\begin{equation}\label{eq:V}
V (z_{i},\Omega_{i})  = 
\left\{
\begin{array}{rl}
\sum_{\alpha} u_{\alpha {\rm S}}w_{\alpha {\rm S}}(z_{i},\Omega_{i}) &{\hspace{1em}\rm if\hspace{1em}} \sigma_{\rm c} <z_{i}<\sigma_{\rm c}+\delta \\
0                          &{\hspace{1em}\rm if\hspace{1em}} z_{i} \ge \sigma_{\rm c}+\delta
\end{array}
\right. 
\end{equation}
where $i$ specifies the particle, $\alpha$ specifies either a patch (P) or the core (C), while $w_{\alpha {\rm S}}$ and $u_{\alpha {\rm S}}$ are the geometric weight factor and the characteristic energy strength of the ${\alpha {\rm S}}$ interaction, respectively. On physical grounds, the particle-substrate interaction range is the same as the particle-particle interaction range.  The energy strengths $u_{\alpha {\rm S}}$ are defined $via$ a mapping between an appropriately developed Debye-H{\"uckel} potential and the model potential~\cite{bianchi:2011}; for system 60n and positive substrate we have $u_{\rm PS}= -0.6278$ and $u_{\rm CS}=2.2310$ (same magnitude but opposite signs for the interactions with the negative substrate), while $u_{\rm PS}=-1.48901$ and $u_{\rm CS}= 2.2310$ for system 60c and positive substrate (again, same magnitude but opposite signs for the interactions with the negative substrate). The weight functions $w_{\alpha {\rm S}}$ are calculated as overlap volumes between the (core or patch) interaction sphere $\alpha$ and the substrate~\cite{bianchi:2d2014}.  For sake of simplicity we omit in the following the explicit dependence of the $w_{\alpha S}$  on $z_{i}$ and $\Omega_{i}$. When $\alpha$ interacts only with one substrate type then the overlap volume is the volume of the spherical cap normalized by the volume of the particle. When $\alpha$ overlaps with multiple stripes we proceed as it follows. First, we compute the overlap cap between $\alpha$ and the substrate as if the substrate were homogeneous: the whole overlap cap is the sum of the overlap sub-volumes represented with dark -- blue and red -- colors in panels (b) and (c) of Figure~\ref{fig:model}. Then, we calculate the volumes of the particle slices each overlap sub-volume belongs to, represented in panel (d) of Figure~\ref{fig:model}. Finally, we use the volumes of the slices to define an effective interaction energy associated to the whole spherical cap, namely $u_{\alpha S}^{*}=({\rm v}_{S1}u_{\alpha S1}+{\rm v}_{S2}u_{\alpha S2})/({\rm v_{S1}+v_{S2}})$, where $\rm v_{S1,S2}$ is the total volume of all particle slices above the substrate of type S1 and S2, respectively. The resulting contribution to the particle-substrate interaction is then $u_{\alpha S}^{*}w_{\alpha {\rm S}}$. We remark that this procedure accounts straightforwardly for an arbitrary large number of interacting stripes. Similar to the particle-particle interaction, also the particle-substrate potential is normalized by the value corresponding to the minimum of the particle-particle attraction $\varepsilon_{\rm min}$.   Characteristic interaction configurations are reported in Fig.~\ref{fig:configs} and are referred to as equatorial-substrate (ES) and polar-substrate (PS).  A representation of the particle-substrate potentials is reported in the Supporting Information of Ref.~\cite{bianchi:2d2014}, while contact energy values for the particle-substrate configurations reported in Fig.~\ref{fig:configs} are listed in Tab.~\ref{tab:contactenergies}.

\begin{table}[h]
\centering
\begin{tabular}{lccccc}
Particle name   & EP & EE & PP & ES$_{(+)}$ & PS$_{(+)}$ \\
\hline
60n &  -1 & 0.159 & 3.683 &  -2.162  & 4.687 \\
60c &  -1 & 0.265 & 0.258 &  -2.611 &  0.347 \\
\end{tabular}
\caption{Contact energies for the investigated IPC-types on a positively charged substrate: two particles, either 60n or 60c, at contact ($r=\sigma$) and fixed mutual orientations, as labeled in Fig.~\ref{fig:configs}. For a graphical representation of the full potentials we refer the reader to Fig.~S2 in the supplementary materials of Ref.~\cite{bianchi:2d2014}. We note that when the substrate is negatively charged, the contact energy values ES and PS just change sign.}
\label{tab:contactenergies}
\end{table}

\subsection{Numerical Methods}
We perform Monte Carlo (MC) simulations of IPC systems with $N=1000$ particles in the canonical ensemble at $T^*=0.10$ (temperature in reduced units,  $i.e.$, $T^*=k_BT/|\varepsilon_{\rm min}|$) in a volume $V=hL^2$, where $L=50\sigma$ and $h=1.45\sigma$. Each MC step consists of $N$ trial particle moves, where the acceptance rule is given by the Metropolis criterion. A particle move is defined as both a displacement in each Cartesian direction of a random quantity distributed uniformly between $\pm \delta r$ as well as a rotation around a random axis of a random angle distributed uniformly between $\pm \delta \theta$. The chosen values for the trial changes are $\delta r = 0.05\sigma$ and  $\delta \theta = \delta r/2\sigma$ rad. We perform for each state point 16 MC parallel runs, starting from different initial conditions. Each run lasts for a total of $\approx  10^8-10^9$ MC steps. All quantities are averages over the final $10^7$ MC steps, corresponding to at least 100 different configurations per independent run. 

\subsection{Post-processing analysis}

To characterize the different aggregation scenarios that emerge on varying the substrate pattern, we (i) study the modulation of the system density (in section~\ref{sec:distr}), (ii) compute the average number of monomers and the average number of neighbors at each state point (in section~\ref{sec:bondorder}) and (iii)  perform a local order analysis using the bond-order parameters $\phi_4$ and $\phi_6$ in two dimensions (in section~\ref{sec:bondorder}). 

Monomers are defined as particles with no bonded interactions, $i.e.,$ the contribution to their energy related to the inter-particle interaction is zero or greater than zero. Neighboring particles are defined following an energy criterion: two particles are considered bonded if their interaction energy is negative. The average number of neighbors is computed accounting only for bonded particles ($i.e.,$ monomers are not counted).
Considering the projection of the system on the $xy$-plane, the bond-order parameters are defined as
\begin{equation}
\phi_4(i) = \left|\frac{1}{\tilde{N}_b(i)} \sum_{j = 0}^{\tilde{N}_b(i)} {e^{i 4 \theta_{ij}}}\right| \qquad \qquad \phi_6(i) = \left|\frac{1}{\tilde{N}_b(i)} \sum_{j = 0}^{\tilde{N}_b(i)} {e^{i 6 \theta_{ij}}}\right|
\label{eq:bondord}
\end{equation}
where $\theta_{ij} = \arctan(y_{ij}/x_{ij})$ is the phase of the relative distance $\boldmath{r}_{ij}$ in two dimensions and $\tilde{N}_b(i)$ is the number of neighbors of particle $i$. 

We also report selected simulation snapshots; for the sake of clarity, we represent the particles with a slightly smaller size than their real one and we use a particle color code to highlight the number of bonds per particle. These two choices help the visualization of compact structures and of their local symmetry.

\section{Results}

\begin{figure}[htbp]
	\begin{center}
	\includegraphics[width=0.5\textwidth]{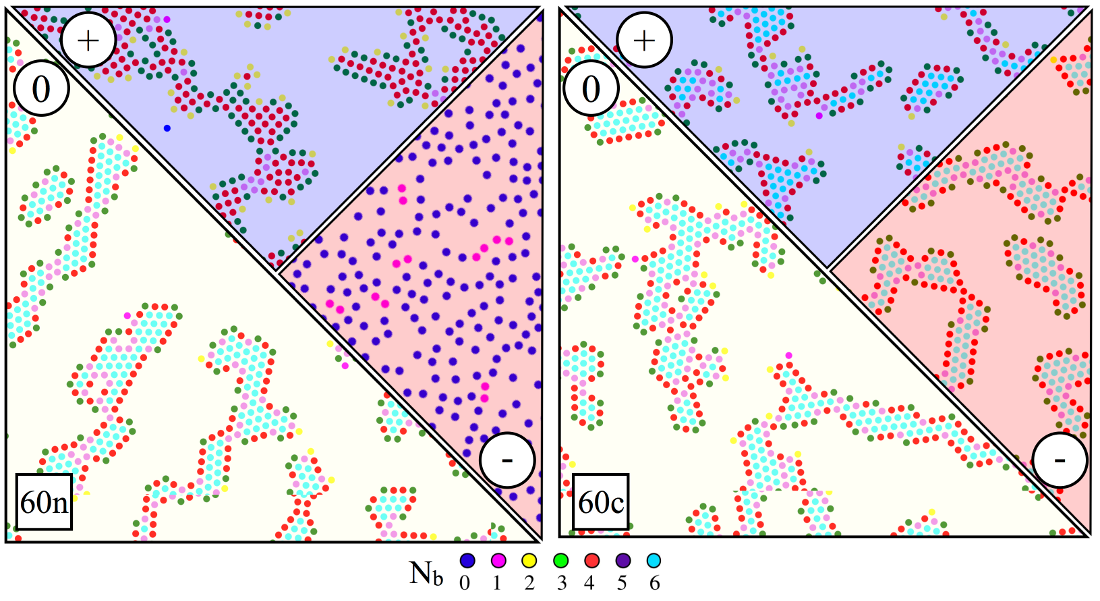} 
	\end{center}
	\caption{Simulation snapshots of IPC systems labeled as 60n (left panel) and 60c (right panel) on a positive (blue), negative (red) and neutral (white) substrate (as labeled). The particle color code refers to the number of bonds per particle, $N_b$: a bond between two particles is formed when their interaction energy is negative.}
	\label{fig:old}
\end{figure}

We start our discussion with a brief summary of the results previously collected for systems 60n and 60c close to a homogeneous substrate~\cite{bianchi:2d2014} (see Figure~\ref{fig:old}). 
When confined between two parallel neutral walls (0) both systems form crystalline domains with a triangular particle arrangement; close to a charged substrate both IPC-types can adsorb on the surface, depending on the sign of the product $Z_{\rm p}Z_{\rm w}$. The competition between the particle-particle and the particle-substrate interactions gives rise to different morphologies. More specifically, a negative substrate (-) disfavors the assembly of system 60n, thus leading to a fluid-like phase where monomeric particles are adsorbed on the substrate; in contrast a positive substrate (+) favors the formation of adsorbed crystalline domains with square-like particle arrangement. For system 60c the triangular arrangement is more or less unaltered~\cite{bianchi:2d2014} with respect to the neutral case, the difference being the adsorption of the domains on the positive substrate, while no adsorption occurs on the negative substrate. 

In the following we describe how the interplay between competing length scales and anisotropic interaction patterns affects the symmetry of the described crystalline domains. 

\subsection{Spatial distribution}
\label{sec:distr}
When the substrate is no longer homogeneous, the particles may prefer a substrate-type with respect to the other and thus occupy specific regions of the sample. To quantify the fluctuations of the particle distribution within the sample, we consider the partial density $\rho_s$, defined as the number of particles observed on a given substrate-type divided by half of the total surface, as a function of $N_s$ for 2~$\le N_s \le $~ 90. We note that, due to the substrate specific design, significant fluctuations in the particle distribution only occur along the $x$ direction (corresponding to the axis of the charge modulation), while along the $y$ direction particles are always uniformly distributed; moreover, since particles are in most cases adsorbed to the charged parts of the substrate,  fluctuations along the $z$ axis are in most cases negligible.
\begin{figure*}[htbp]
	\begin{center}
	\includegraphics[width=0.75\textwidth]{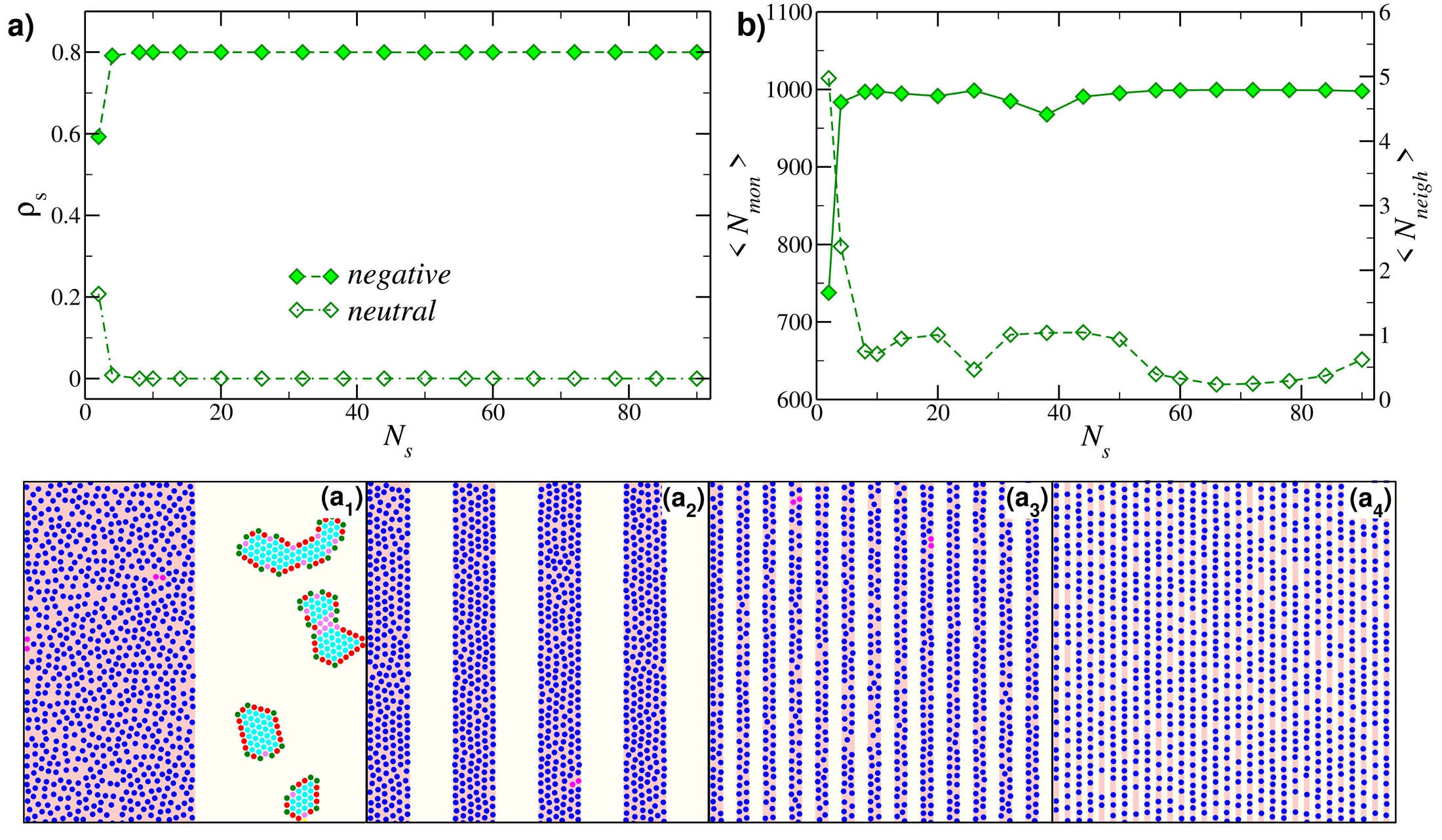} 
	\end{center}
	\caption{System 60n on a -/0 substrate. Upper panels: a) partial densities $\rho_{\rm s}$ on the two substrate types (as labelled) as a function of the number of stripes $N_s$, b) average number of monomers $\langle N_{mon} \rangle$ (full symbols) and average number of neighbors $\langle N_{neigh} \rangle$ (empty symbols) as a function of $N_s$. Lower panels: MC simulations snapshots at selected state points: ($a_1$) $N_s =$ 2, ($a_2$) $N_s =$ 8, ($a_3$) $N_s =$ 26 and ($a_4$) $N_s =$ 60. The particle color code refers to the number of bonds per particle and is it reported in Figure~\ref{fig:old}.}
	\label{fig:60n_n0}
\end{figure*}

We start our discussion with system 60n. On a -/0 substrate with $N_s=2$, particles form crystalline, non-adsorbed domains on the neutral stripe and adsorb on the negative stripe without forming any aggregate (see snapshot ($\rm a_1$) in Fig.~\ref{fig:60n_n0}). As soon as $N_s$ increases, particles prefer to absorb on the negatively charged regions, thus leaving the neutral stripes completely empty (see snapshots ($\rm a_2$), ($\rm a_3$) and ($\rm a_4$) in Fig.~\ref{fig:60n_n0}). The corresponding $\rho_s$ (see panel a) of Fig.~\ref{fig:60n_n0})  clearly confirms that for $N_s=2$ most particles are located on the negative stripes and for $N_s\ge8$ the neutral stripes are completely empty. All populated stripes are characterized by adsorbed monomers.

\begin{figure*}[htbp]
	\begin{center}
	\includegraphics[width=0.75\textwidth]{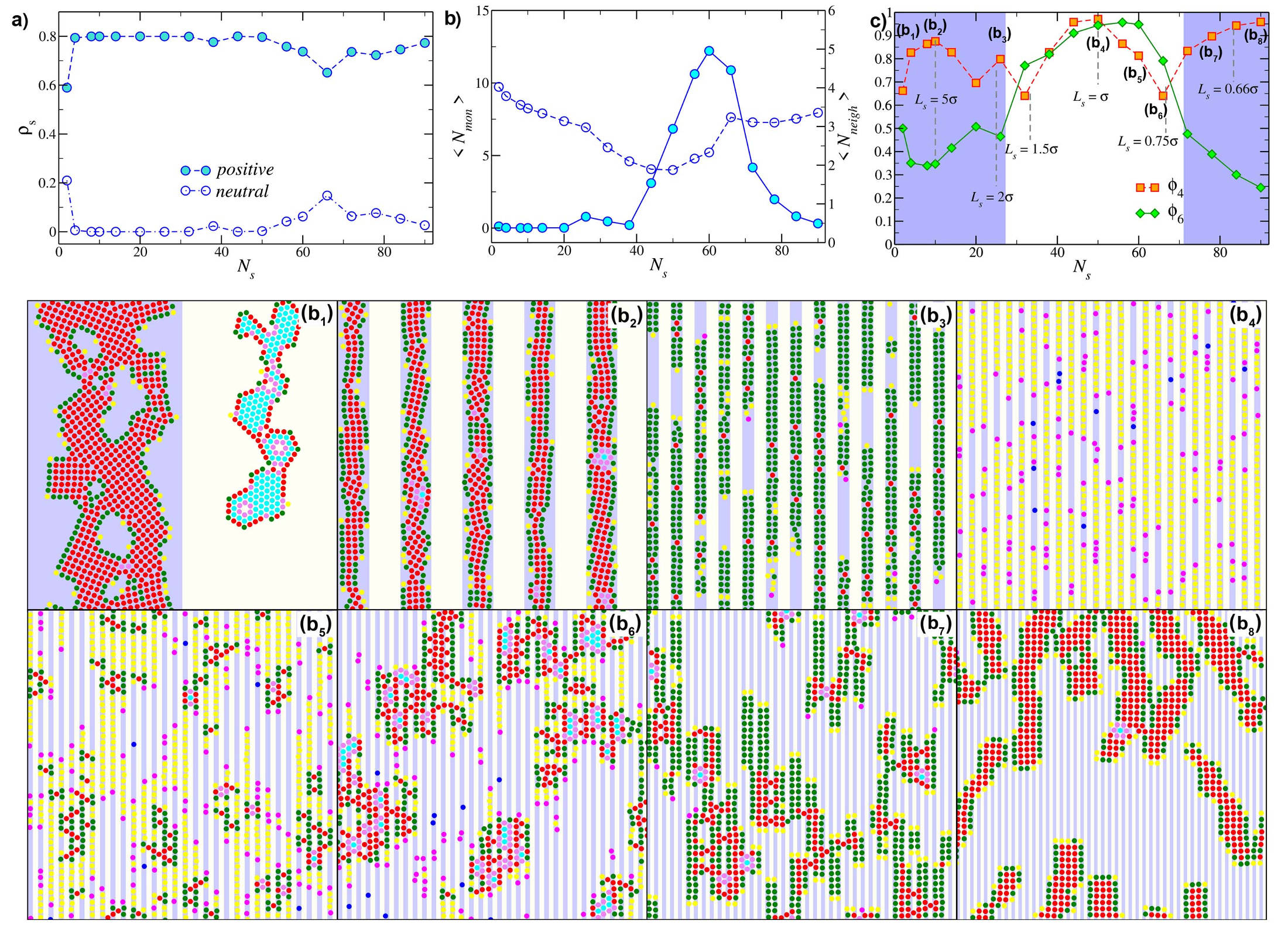} 
	\end{center}
	\caption{System 60n on a +/0 substrate. Upper panels: a) partial densities $\rho_{\rm s}$ on the two substrate types (as labelled) as a function of the number of stripes $N_s$, b) average number of monomers $\langle N_{mon} \rangle$ (full symbols) and average number of neighbors $\langle N_{neigh} \rangle$ (empty symbols) as a function of $N_s$, c) local bond order parameters $\phi_4$ and $\phi_6$ as function of $N_s$, shaded sectors highlight regions where the statistics is relevant. Lower panels: MC simulations snapshots at selected state points: ($b_1$) $N_s =$ 2, ($b_2$) $N_s =$ 10, ($b_3$) $N_s =$ 26, ($b_4$) $N_s =$ 50, ($b_5$) $N_s =$ 60, ($b_6$) $N_s =$ 66, ($b_7$) $N_s =$ 78 and ($b_8$) $N_s =$ 90. The particle color code refers to the number of bonds per particle and is it reported in Figure~\ref{fig:old}.}
	\label{fig:60n_p0}
\end{figure*}

The neutral stripes completely deplete -- as soon as $N_s >$ 2 -- also for system 60n on a +/0 substrate (see Fig.~\ref{fig:60n_p0}). In contrast to the previous case, particles adsorbed on the charged stripes form large crystalline clusters. By visual inspection, we can anticipate that the crystalline clusters become more and more elongated with increasing $N_s$ (see snapshots ($\rm b_1$) and ($\rm b_2$) in Fig.~\ref{fig:60n_p0}); on further increasing $N_s$ particles organize in double-particle lanes (see snapshot ($\rm b_3$) in Fig.~\ref{fig:60n_p0}); when the size of the stripes becomes comparable to $\sigma$ ($i.e.$, for $N_s = 50$) particles organize in single-particle strings (see snapshot ($\rm b_4$) in Fig.~\ref{fig:60n_p0}); finally, these strings merge and form again a crystalline pattern  (see snapshot ($\rm b_8$) in Fig.~\ref{fig:60n_p0}). The symmetry of the emerging aggregates depends on the competition between the different energies and length scales and it is discussed in section~\ref{sec:bondorder}.  

\begin{figure*}[htbp]
	\begin{center}
	\includegraphics[width=0.75\textwidth]{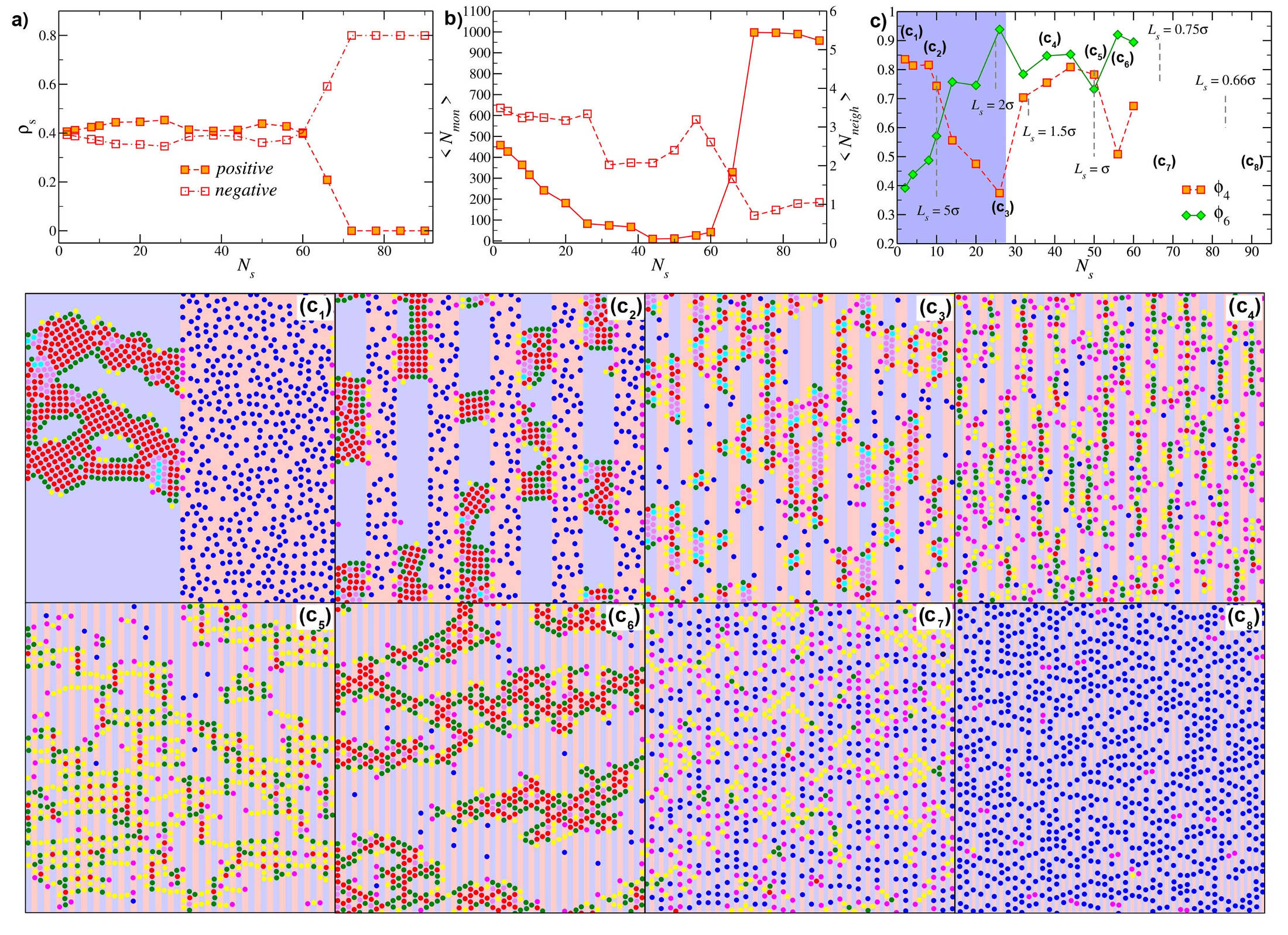} 
	\end{center}
	\caption{System 60n on a +/- substrate. Upper panels: a) partial densities $\rho_{\rm s}$ on the two substrate types (as labelled) as a function of the number of stripes $N_s$, b) average number of monomers $\langle N_{mon} \rangle$ (full symbols) and average number of neighbors $\langle N_{neigh} \rangle$ (empty symbols) as a function of $N_s$, c) local bond order parameters $\phi_4$ and $\phi_6$ as function of $N_s$,  shaded sectors highlight regions where the statistics is relevant. Lower panels: MC simulations snapshots at selected state points: ($c_1$) $N_s =$ 2, ($c_2$) $N_s =$ 10, ($c_3$) $N_s =$ 26, ($c_4$) $N_s =$ 38, ($c_5$) $N_s =$ 50, ($c_6$) $N_s =$ 56, ($c_7$) $N_s =$ 66 and ($c_8$) $N_s =$ 90. The particle color code refers to the number of bonds per particle and is it reported in Figure~\ref{fig:old}.}
	\label{fig:60n_opposite}
\end{figure*}

When particles adsorb on both substrate-types the scenario becomes more complex. Results for system 60n on a +/- substrate are shown in Fig.~\ref{fig:60n_opposite}. Even though the attraction between the patches and the negative substrate is stronger than the attraction between the equators and the positive substrates (see Tab.~\ref{tab:contactenergies}), as long as $N_s\le60$, particles slightly prefer to adsorb on the positive stripes ($i.e.$, $\rho_s$-values for the positive and negative stripes are comparable to the average system density $\rho=0.4$, see panel a) of Fig.~\ref{fig:60n_opposite}), where they can also aggregate. Visually, clusters on the positive stripes elongate and reshape on increasing $N_s$ (see snapshots ($c_1$)-($c_4$) in Fig.~\ref{fig:60n_opposite}) to the point that, when the width of the stripes is equal to $\sigma$ (and slightly smaller), particles form extended aggregates spanning across several adjacent stripes (see snapshot ($c_5$) and ($c_6$) in Fig.~\ref{fig:60n_opposite}). The transition from vertical aggregates to extended clusters is discussed in section~\ref{sec:bondorder} and~\ref{sec:visual}. Finally, when $N_s > 60$, particles prefer to adsorb on the negative stripes, as signaled by the abrupt increase of the partial density associated to the negative stripes towards $2\rho=0.8$  (see panel a) in Fig.~\ref{fig:60n_opposite}). Particles adsorbed on the negative stripes do not bond to each other (see snapshot ($c_7$) and ($c_8$) in Fig.~\ref{fig:60n_opposite}). 

In summary, for system 60n, $\rho_s$ depicts the following scenario: for patterns -/0 and +/0, the neutral stripes are completely empty already at very low $N_s$-values; for pattern +/- the partial densities of the two substrate-types are very similar to each other, with a slight preference for the positive substrate as long as $N_s$ is small enough, while at high $N_s$-values, namely, when the width of the stripes is smaller than $\sigma$, the negative stripes become favorite. Anticipating a more detailed analysis proposed in section~\ref{sec:bondorder} and~\ref{sec:visual}, we also observe that, when particles are preferentially adsorbed to the negative stripes, no aggregation occurs at $N_s=90$ (see Fig.~\ref{fig:60n_n0} and Fig.~\ref{fig:60n_opposite}), while, when particles adsorb on the positive stripes (see Fig.~\ref{fig:60n_p0}) a crystalline pattern is formed at $N_s=90$.  

\begin{figure*}[htbp]
	\begin{center} 
		\includegraphics[width=0.75\textwidth]{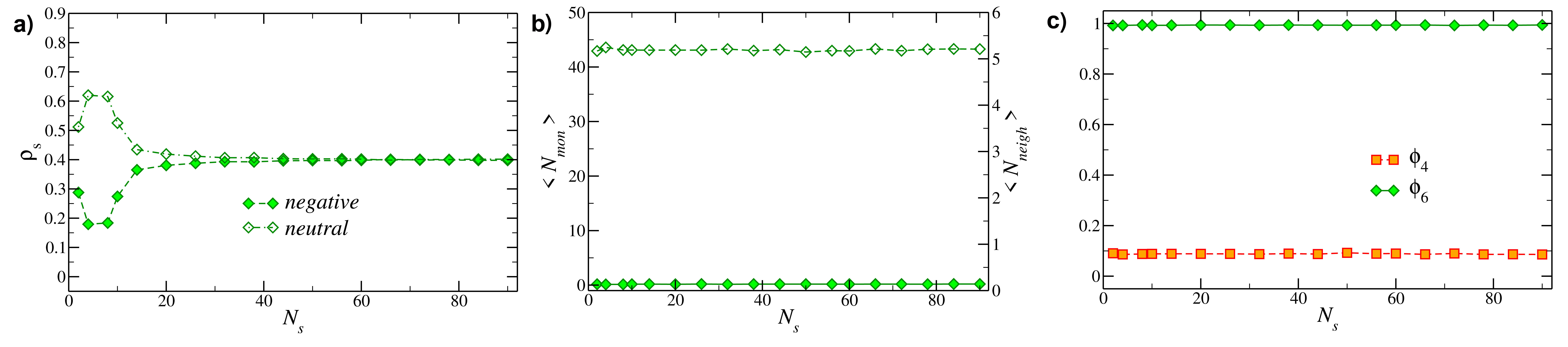}
	\end{center}
	\caption{System 60c on a -/0 substrate. Panel a) partial densities $\rho_{\rm s}$ on the two substrate types (as labelled) as a function of the number of stripes $N_s$. Panel b) average number of monomers $\langle N_{mon} \rangle$ (full symbols) and average number of neighbors $\langle N_{neigh} \rangle$ (empty symbols) as a function of $N_s$. Panel c) local bond order parameters $\phi_4$ and $\phi_6$ as function of $N_s$.}
	\label{fig:60c_n0}
\end{figure*}

We now look at the corresponding data for systems 60c. On a -/0 substrate, the system forms crystalline domains almost irrespective of the substrate charge, since particles do not adsorb on the neutral nor on the negative stripes~\cite{bianchi:2d2014}. The analysis of the partial densities shows a slight preference for the neutral substrate at small $N_s$-values, which soon disappears (see panel a) of Fig.~\ref{fig:60c_n0}): already for $N_s \approx 20$ the partial densities of both substrate-types reach the average density of the system $\rho=0.4$. 

\begin{figure*}[htbp]
	\begin{center}
	\includegraphics[width=0.75\textwidth]{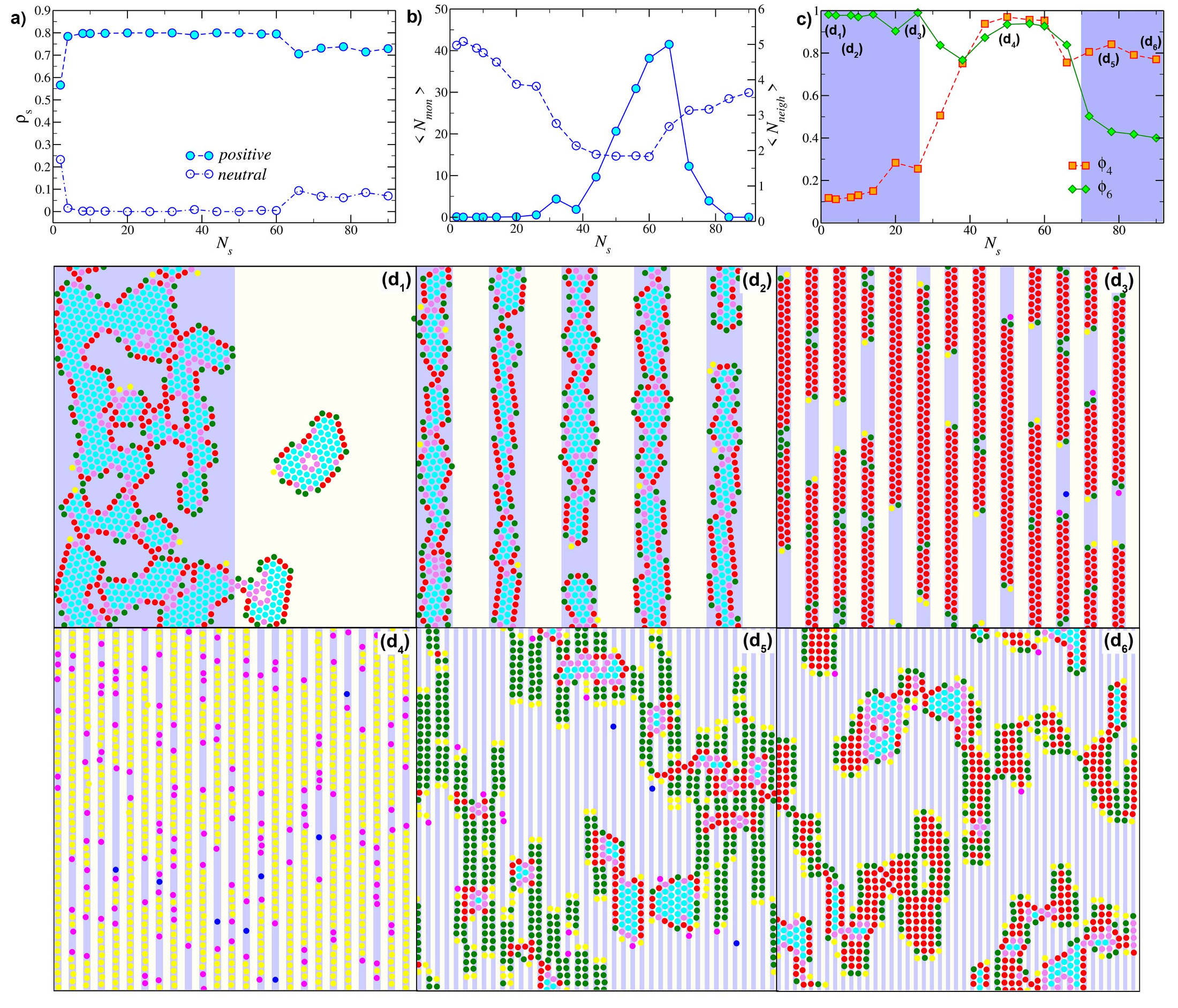}
	\end{center}
	\caption{System 60c on a +/0 substrate. Upper panels: a) partial densities $\rho_{\rm s}$ on the two substrate types (as labelled) as a function of the number of stripes $N_s$, b) average number of monomers $\langle N_{mon} \rangle$ (full symbols) and average number of neighbors $\langle N_{neigh} \rangle$ (empty symbols) as a function of $N_s$, c) local bond order parameters $\phi_4$ and $\phi_6$ as function of $N_s$,  shaded sectors highlight regions where the statistics is relevant. Lower panels: MC simulations snapshots at selected state points: ($d_1$) $N_s =$ 2, ($d_2$) $N_s =$ 10, ($d_3$) $N_s =$ 26, ($d_4$) $N_s =$ 50, ($d_5$) $N_s =$ 78 and ($d_6$) $N_s =$ 90. The particle color code refers to the number of bonds per particle and is it reported in Figure~\ref{fig:old}.}
	\label{fig:60c_p0}
\end{figure*}

Similar to the corresponding case for the 60n system, on a +/0 substrate, particles prefer to assemble on the positive stripes where they also adsorb. As soon as $N_s > 2$ (see panel a) of Fig.~\ref{fig:60c_p0}), the neutral stripes become empty and all particles adsorb on the positive stripes forming crystalline domains ($i.e.$, $\rho_s=0.8$ on the positive stripes); these domains become more and more elongated and narrow on increasing $N_s$; as soon as the width of the stripes is smaller than $\sigma$, particles organize in single-particle strings that eventually merge into extended clusters, as discussed in section~\ref{sec:bondorder}. We observe that, for $N_s>60$, $\rho_s$ on the positive (neutral) substrate slightly decreases (increases): a visual inspection of the configurations suggests that the emergent crystalline domains have distinct symmetries, as discussed in section~\ref{sec:bondorder}. 

\begin{figure*}[htbp]
	\begin{center}
	\includegraphics[width=0.75\textwidth]{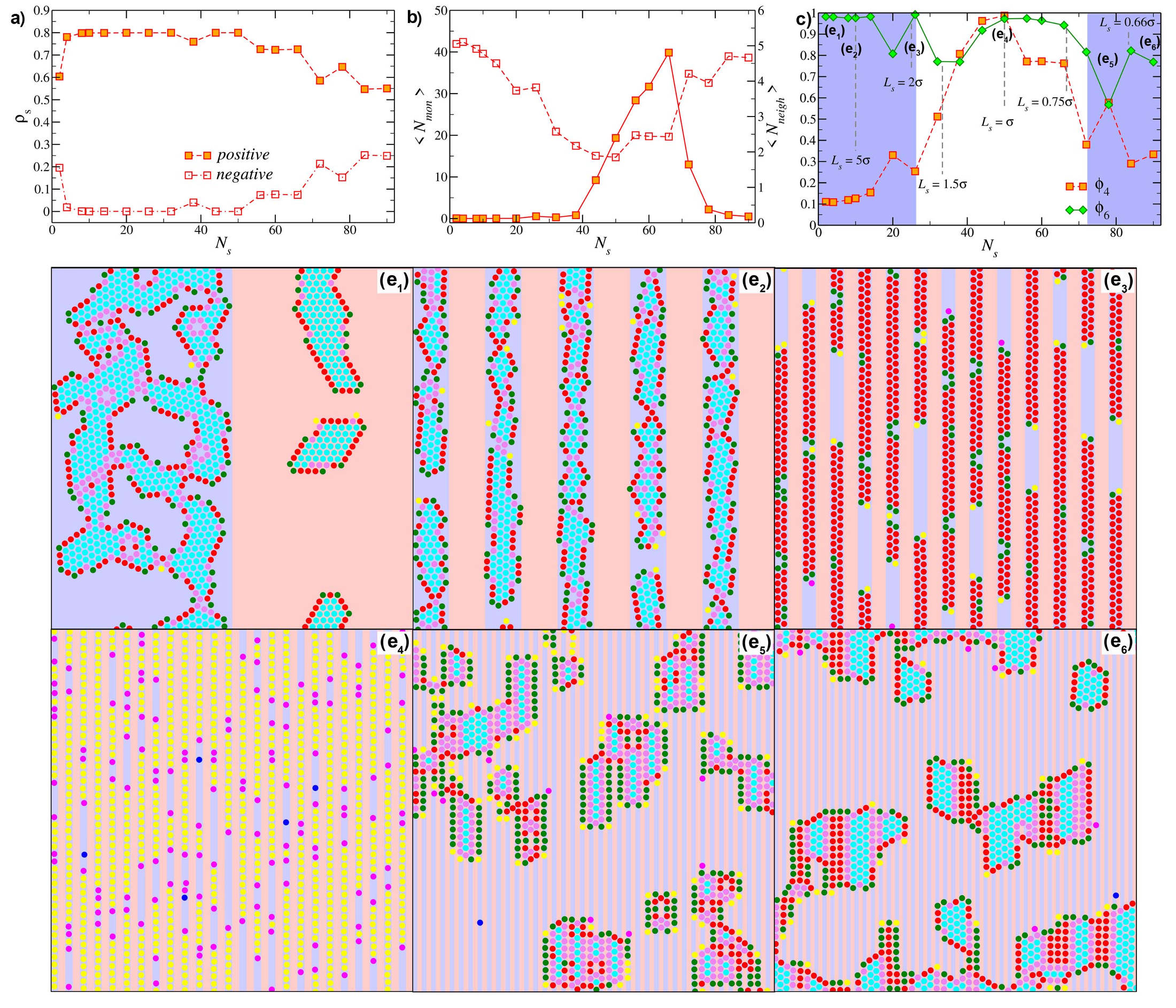} 
	\end{center}
	\caption{System 60c on a +/- substrate. Upper panels: a) partial densities $\rho_{\rm s}$ on the two substrate types (as labelled) as a function of the number of stripes $N_s$, b) average number of monomers $\langle N_{mon} \rangle$ (full symbols) and average number of neighbors $\langle N_{neigh} \rangle$ (empty symbols) as a function of $N_s$, c) local bond order parameters $\phi_4$ and $\phi_6$ as function of $N_s$,  shaded sectors highlight regions where the statistics is relevant. Lower panels: MC simulations snapshots at selected state points: ($e_1$) $N_s =$ 2, ($e_2$) $N_s =$ 10, ($e_3$) $N_s =$ 26, ($e_4$) $N_s =$ 50, ($e_5$) $N_s =$ 78 and ($e_6$) $N_s =$ 90. The particle color code refers to the number of bonds per particle and is it reported in Figure~\ref{fig:old}.}
	\label{fig:60c_opposite}
\end{figure*}

On a +/- substrate, 60c particles prefer again to  occupy the positive stripes (since 60c particles do not adsorb on the negative substrate~\cite{bianchi:2d2014}) and the assembly scenario is similar to the +/0 case (see panel a) of Fig.~\ref{fig:60c_opposite}). In this case, when $N_s>60$, the decrease (increase) of $\rho_s$ on the positive (negative) substrate is bigger than in the previous case, and this is  related to the symmetry of the aggregates, as discussed in section~\ref{sec:bondorder}.


In summary, for system 60c, $\rho_s$ depicts the following scenario:  for pattern -/0 the system is almost insensitive to the presence of the stripes on the substrate, while for patterns +/0 and +/-, the positive stripes are mostly preferred and the particle assembly ranges from extended aggregates at small $N_s$-values, to elongated clusters, double- and single-particle lanes and finally crystalline domains again. 

It is worth stressing that, for both systems, as soon as the particles prefer to adsorb to the positive stripes, we observe the emergence of crystalline domains at $N_s=90$: the features of these domains differ from case to case since they depend on the delicate balance between different factors. In the following we perform a neighbor analysis in order to describe the different aggregation scenarios sketched above.

\subsection{Neighbor analysis}
\label{sec:bondorder}

To characterize the symmetry of the domains formed by IPCs in presence of a patterned substrate we consider the average number of monomers $\langle N_{mon} \rangle$, the average number of neighbors $\langle N_{neigh} \rangle$, and the local bond order parameters $\phi_4$ and $\phi_6$. We have highlighted in blue, in all panels c) of Figs.~\ref{fig:60n_p0},~\ref{fig:60n_opposite},~\ref{fig:60c_p0} and~\ref{fig:60c_opposite}, the state points where clusters with crystalline arrangements have been observed. In the other regions, non-crystalline arrangements occur and the values of the bond order parameters should not be used as an indicator of the assembly state of the system.
 
We start again our discussion with system 60n. As mentioned in section~\ref{sec:distr}, when system 60n is close to a -/0 pattern most of the particles are monomers adsorbed to the substrate (see panel b) of Fig.~\ref{fig:60n_n0}). More precisely, at $N_s=2$, slightly more than half of the sample is in a monomer state (adsorbed to the negative stripe), while the remaining particles belong to the crystalline domains (formed in correspondence to the neutral stripe).  We know from the homogeneous case (see Fig.~\ref{fig:old}, left panel) that the crystalline domains observed on a neutral substrate have a triangular arrangement~\cite{bianchi:2d2014} and this is consistent with the $\langle N_{neigh} \rangle$-value observed at $N_s=2$.  As soon as $N_s>2$, $\langle N_{mon} \rangle$ rapidly levels off to $N$, $i.e.,$ most of the particles become monomers. This is consistent with the corresponding drop of $\langle N_{neigh} \rangle$ from five to zero. In this case, a bond order analysis has little significance (and thus it is not reported), due to poor statistics. 

When we consider a +/0 pattern, we observe that $\langle N_{mon} \rangle$ is negligible at all $N_s$-values (see panel b) of Fig.~\ref{fig:60n_p0}), since its peak for $40<N_s< 60$ corresponds to $\langle N_{mon} \rangle \approx 13$. In this case, particles form crystalline domains (see panel ($b_1$) of Fig.~\ref{fig:60n_p0}) that get elongated on increasing $N_s$, from compact clusters (see panel ($b_2$) of Fig.~\ref{fig:60n_p0}) to two-particles lanes (see panel ($b_3$) of Fig.~\ref{fig:60n_p0}) and single-particle strings (see panel ($b_4$) of Fig.~\ref{fig:60n_p0}), that eventually merge into crystalline domains again (see panel ($b_5$)-($b_8$) of Fig.~\ref{fig:60n_p0}), as anticipated in section~\ref{sec:distr}. The corresponding $\langle N_{neigh} \rangle$  (see panel b) of Fig.~\ref{fig:60n_p0}) decreases from  $\langle N_{neigh} \rangle=4$ at $N_s=2$ (compact domains) to  $\langle N_{neigh} \rangle=2$ at $N_s=50$ (single-particle stripes) and increases again to reach  $\langle N_{neigh} \rangle=3$ at $N_s=90$  (compact domains). To better characterize the symmetries of the emerging compact domains we consider $\phi_4$ and $\phi_6$ as function of $N_s$. In panel c) of Fig.~\ref{fig:60n_p0} the two regimes where the order parameters are relevant are highlighted in blue. Both regimes correspond to a predominance of square particle arrangements: at small $N_s$  ($N_s<40$, panels ($b_1$)-($b_3$) of Fig.~\ref{fig:60n_p0}) this is the effect of the competition between different anisotropic interactions, while at large $N_s$ ($N_s> 72$, panels ($b_7$) and ($b_8$) of Fig.~\ref{fig:60n_p0}) an additional competition between different length scales sets in. More specifically, at small $N_s$, after the initial coexistence between square and triangular lattices (see snapshot ($b_1$) in Fig.~\ref{fig:60n_n0}), neutral stripes get depleted and the system forms clusters adsorbed on the positive stripes with a square symmetry, as signaled by the increase (decrease) of $\phi_4$ ($\phi_6$) for $N_s>2$. Despite becoming narrower and more elongated on increasing $N_s$, clusters maintain a square-like order until $N_s \simeq$ 32 (see snapshots ($b_2$) and ($b_3$) in Fig.~\ref{fig:60n_p0}), where particle organize in square-like, double-particle lanes. On further increasing $N_s$, we observe a coexistence between double-particle lanes with square symmetry and single-particle strings, which eventually prevail as soon as the particles are not able to pair in line on the same stripe anymore, $i. e.$, at $N_s =$ 50 (see snapshot ($b_4$) in Fig.~\ref{fig:60n_n0}). On further decreasing the size of the stripes below $\sigma$, these single-particle strings start merging (see snapshot ($b_5$) in Fig.~\ref{fig:60n_n0}):  the effect of the stripes is to create \emph{tracks} to guide the assembly; over these tracks, particles are arranged in a configuration that greatly favors square domains. Thus, even tough there is a sweet spot where triangular, packed arrangements emerge (see snapshot ($b_6$) in Fig.~\ref{fig:60n_n0}), upon increasing $N_s$ single-particle stripes are just led to merge together (see snapshot ($b_7$) in Fig.~\ref{fig:60n_n0}), to finally form square domains (see snapshot ($b_8$) in Fig.~\ref{fig:60n_n0}), as signaled by the abrupt drop (growth) of $\phi_6$ ($\phi_4$).  In such a regime, we see clearly the onset of a strong pattern-induced particle arrangement, which span over large areas of the substrate itself. We stress that at all state points the crystalline arrangements are realized with particles mostly adsorbed on the positive stripes, as discussed in section~\ref{sec:distr}.

Finally, for system 60n close to a +/- substrate, at small $N_s$ slightly less than half of the sample is in a monomer state ($i.e.$, $\langle N_{mon} \rangle \approx 500$), while the remaining particles belong to the crystalline domains formed in correspondence to the positive stripes. At low $N_s$-values, large $\phi_4$-values are observed, suggesting that the particles in crystalline clusters have a square-like symmetry (see snapshots ($c_1$) and ($c_2$) in Fig.~\ref{fig:60n_opposite}).  When $N_s \gtrsim10$, square domains break down in favor of double-particle lanes with triangular symmetry -- again on the positive stripes -- and $\phi_4 < \phi_6$ (see snapshots ($c_3$) of Fig.~\ref{fig:60n_opposite}). When $N_s \approx 26$, one single stripe cannot accomodate two fully embedded particles anymore, so single-particle strings form on the positive stripes, while the monomers adsorbed on the negative stripes act as boundaries of these lanes (see snapshot ($c_4$) of Fig.~\ref{fig:60n_opposite}). When $40<N_s< 70$,  $\langle N_{mon} \rangle$ drops to zero. Within this region, as long as  $N_s \gtrsim50$, single-particle strings on the positive stripes becomes connected $via$ monomers on the negative stripes and a visual inspection of the particle configurations suggests a very rich scenario: networks emerge with properties of that are deeply affected by the substrate pattern.  In this case, the behavior of $\langle N_{neigh} \rangle$ is less significative. Additionally, since clusters in this $N_s$-region are not compact anymore, the local bond order parameters are not reliable and a different analysis, which is beyond the scope of this paper, is needed. Finally, for $N_s\gtrsim70$  monomers prevail as shown by the abrupt increase of $\langle N_{mon} \rangle$ (see panel c) of Fig.~\ref{fig:60n_opposite}).

To summarize, system 60n shows a very rich assembly scenario, where according to the charge modulation it is possible to induce the formation of (square-like or triangular-like) double-particle lanes, open networks spanning throughout the sample and pattern-induced square domains.  

We now consider system 60c. As anticipated in section~\ref{sec:distr}, for the -/0 pattern particles tend to cluster, irrespective of the substrate type.
The analysis of $\langle N_{mon} \rangle$ and $\langle N_{neigh} \rangle$ suggests that only one crystalline symmetry, the triangular one, is maintained over the whole $N_s$ range: at all $N_s$ values, $\langle N_{mon} \rangle = 0$  and $\langle N_{neigh} \rangle \simeq 5$, suggesting the presence of clusters with a triangular particle arrangement (see panel b) Fig.~\ref{fig:60c_n0}). The relative local bond oder parameters support this scenario, since $\phi_6 \simeq 1$ and $\phi_4 \simeq 0$ at any $N_s$-value (see panel c) of Fig.~\ref{fig:60c_n0}). 

In contrast, the +/0 and the +/- patterns show a very similar trend, corresponding to a transition from a regime of purely triangular domains to a regime where triangular-like domains coexist with square-like aggregates. The similarity between the two cases is due to the fact that particles 60c strongly absorb on the positive substrate, that is thus preferred to both the negative and neutral stripes. The transition from one crystalline regime to another can be inferred already from the analysis of  $\langle N_{mon} \rangle$ and $\langle N_{neigh} \rangle$ (see panels b) of Fig.s~\ref{fig:60c_p0} and~\ref{fig:60c_opposite}): at low $N_s$ no monomers are found (for both substrate patterns) and particles have on average five neighbors (suggesting a predominance of triangular domains); at large $N_s$ very few monomers are observed and particles have on average 3.5 (+/0 substrate) and 4.5 (+/- substrate) neighbors, suggesting the presence of clusters with mixed symmetry. Between these two regimes $\langle N_{neigh} \rangle$ drops to 2 (together with a mild increase of $\langle N_{mon} \rangle$ to 40), suggesting that the crystalline order is broken and restored.  The transition between the two regimes is similar to the one observed for system 60n close to a +/0 pattern: as long as the stripes are sufficiently large, particles form extended crystalline clusters; on increasing $N_s$, particles start to organize in double-particle lanes on the same stripe (see snapshots ($\rm d_3$) of Fig.~\ref{fig:60c_p0} and ($\rm e_3$) of Fig.~\ref{fig:60c_opposite}), then  in single-particle strings (see snapshots ($\rm d_4$) of Fig.~\ref{fig:60c_p0} and ($\rm e_4$) of Fig.~\ref{fig:60c_opposite}) and finally at very large $N_s$ these strings form big domains (see snapshots ($\rm d_6$) of Fig.~\ref{fig:60c_p0} and ($\rm e_6$) of Fig.~\ref{fig:60c_opposite}). From the behavior of the local bond order parameters as function of $N_s$, we observe that at $N_s$=90, square-like domains prevail ($\phi_4 \approx 0.4$ while $\phi_6 \approx 0.8$) in the +/0 case, while triangular domains prevail ($\phi_4 \approx 0.8$ while $\phi_6 \approx 0.4$) in the +/- case. The competition between square- and triangular-arrangements is directly related to the balance between energy gains and losses in the particle-substrate and particle-particle interaction: while square domains favor the particle adsorption, triangular domains favor inter-particle bonding. Here, particles do not pay any energetic price when oriented in-plane on a neutral substrate, while the would pay a price for such an orientation on a negative stripe; for this reason more triangular domains are observed on a +/- substrate with respect to the +/0  case. 

We observe that, while for system 60n on a +/- substrate the emergence of square domains at high $N_s$-values is compatibile with the behavior of the system at low $N_s$-values  (where square domains prevail), the behavior of system 60c at low $N_s$-values (where triangular domains prevail) is quite different from the behavior at high $N_s$-values (where pattern-induced square domains emerge). 
Finally, we note that, with respect to system 60n, systems 60c does not form open networks nor square-like double-particle lanes. 

\subsection{Robustness of the self-assembly scenario}
\label{sec:var}

We discuss here the robustness of the assembly scenarios presented in sections~\ref{sec:distr} and~\ref{sec:bondorder} by considering small variations of the system parameters. In particular, we investigate the effects of different (i) surface charges and (ii) interaction ranges, the latter being related to the electrostatic screening conditions. 

Since our model is not a toy model but a coarse-grained description of the effective interactions, the interaction parameters must be computer anew $via$ the procedure described in section~\ref{sec:model&methods} and reported in more details in Ref.~\cite{bianchi:2011}. The list of parameters considered here is reported in Tab.~\ref{tab:addparams}: for all the systems, $\cos \gamma \sim 60 \degree$, the geometric parameters follows Eq.~(\ref{eq:params}) and the energy constants are calculated following the mapping procedure introduced in Ref.~\cite{bianchi:2011}.

\begin{table*}[htb]
\centering
\begin{tabular}{lccccccccc}
Particle name   & $Z_w$ & $\delta$ & $a$ & $u_{CC}$ & $u_{PC}$ & $u_{PP}$ & $u_{CS_{(+)}}$ & $u_{PS_{(+)}}$ & $\varepsilon_{min}$ \\
\hline
60n & 1.0 & 0.20 & 0.1600 & 0.1349  & -0.8483  & 4.3228  & -0.1256 & 0.4462 & -0.0337   \\
60n & 1.5 & 0.20 & 0.1600 & 0.1349  & -0.8483  & 4.3228  & -0.1883 & 0.6693 & -0.0337  \\
60n & 3.0 & 0.20 &  0.1600 & 0.1349 & -0.8483  & 4.3228  & -0.3767 & 1.3385 & -0.0337   \\
60n & 5.0 & 0.20 &  0.1600 & 0.1349 & -0.8483 & 4.3228   & -0.6278 &  2.2310 & -0.0337  \\
60c & 1.0 & 0.20 &  0.1600 & 0.4330 & -1.9467  & 4.3228  & -0.2978 & 0.4462 & -0.0781   \\
60c & 1.5 & 0.20 & 0.1600 & 0.4330  & -1.9467  & 4.3228  & -0.4467 & 0.6693 & -0.0781   \\
60c & 3.0 & 0.20 &  0.1600 & 0.4330 & -1.9467  & 4.3228  & -0.8934 & 1.3386 & -0.0781   \\
60c & 5.0 & 0.20 &  0.1600 & 0.4330 & -1.9467  & 4.3228  & -1.4890 & 2.2310 & -0.0781  \\
\hline
SR & 5.0 & 0.15 &  0.1265 & 0.0798 & -0.5050  & 2.5470 & -0.6694 &  2.3663 & -0.0123  \\
LR & 5.0 & 0.25 & 0.1907 & 0.2160 & -1.3068  & 6.6472 & -0.6019 &  2.1261 & -0.0736  \\
\end{tabular}
\caption{List of the model parameters used to induce variations either in the particle-substrate interaction (upper part) or in the particle-particle interaction (lower part): substrate charge, $Z_w$,  interaction range, $\delta$,  eccentricity $a$, energy parameter for the core/core (CC), patch-core (PC), patch-patch (PP), core-substrate (CS) and patch-substrate (PS) interaction (in case of positive substrate) and minimum of the attraction, $\varepsilon_{min}$.}
\label{tab:addparams}
\end{table*}

\subsubsection{Effect of $Z_w$}

\begin{figure*}[t]
	\begin{center}
	\includegraphics[width=0.75\textwidth]{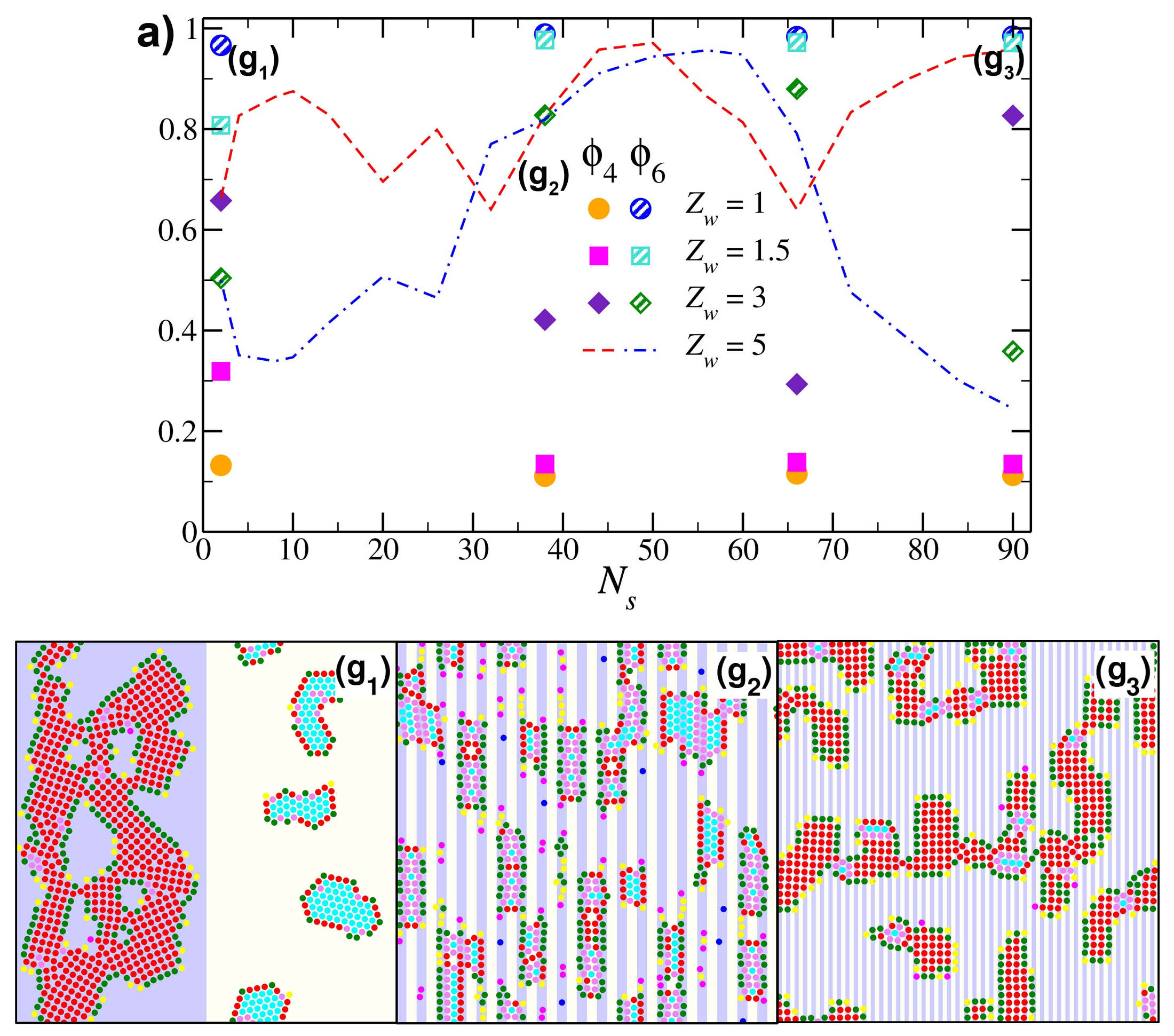} 
	\end{center}
	\caption{System 60n substrate +/0: local bond order parameters, $\phi_4$f and $\phi_6$, as function of $N_s$ for different $Z_w$-values (as labeled): symbols refer to $Z_w =$ 1, 1.5, 3, while lines refer to $Z_w =$ 5 (reported as a reference). Data are reported for selected $N_s$, namely $N_s=2,38,66$, and 90. Panels ($g_1$)-($g_3$) are snapshots of the system with $Z_w =$ 3 at different number of stripes: ($g_1$) $N_s =$ 2, ($g_2$) $N_s =$ 66, ($g_3$) $N_s =$ 90. }
	\label{fig:surfcharge_n1}
\end{figure*}

\begin{figure*}[t]
	\begin{center}
	\includegraphics[width=0.75\textwidth]{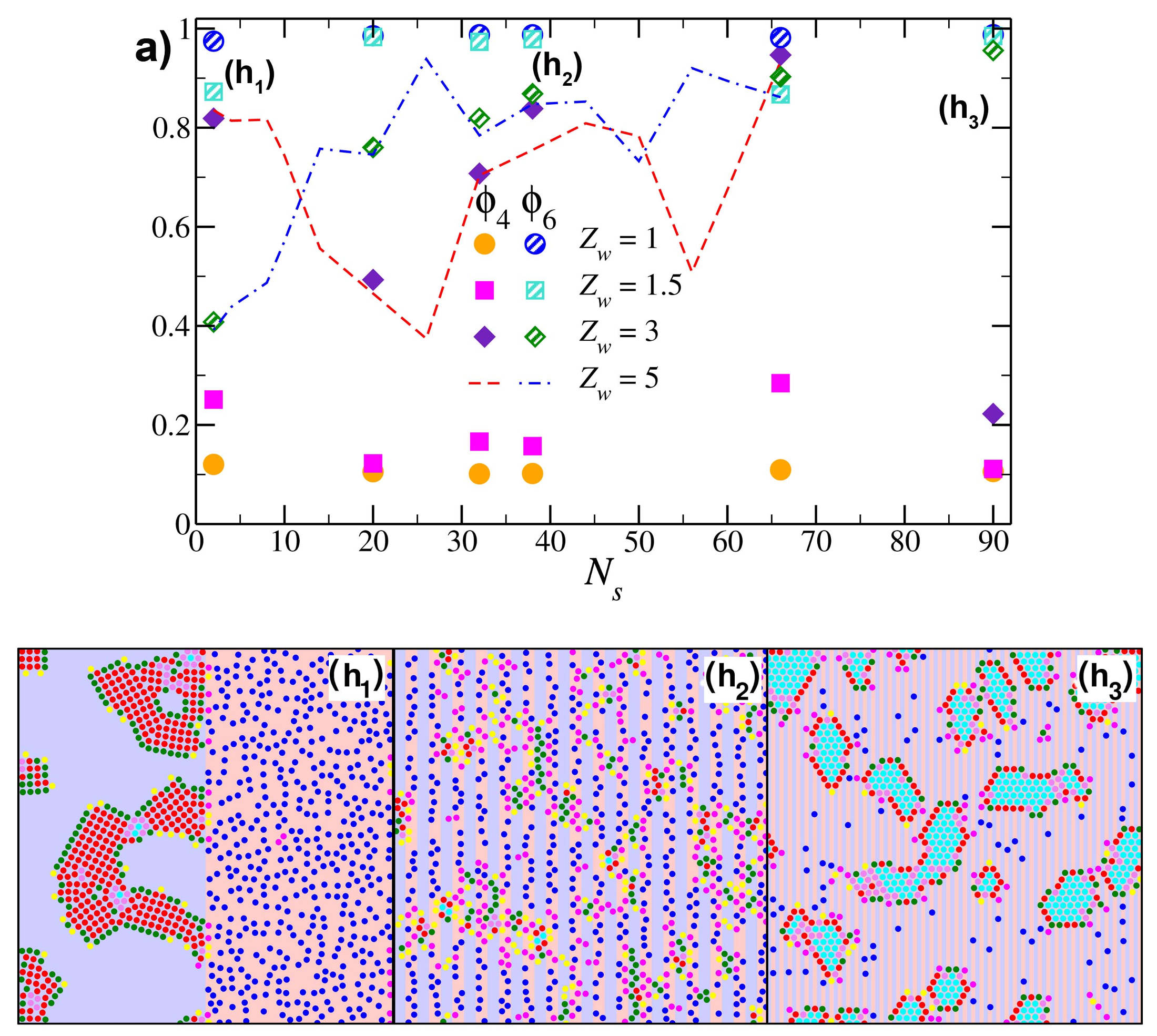} 
	\end{center}
	\caption{System 60n substrate +/-: local bond order parameters, $\phi_4$f and $\phi_6$, as function of $N_s$ for different $Z_w$-values (as labeled): symbols refer to $Z_w =$ 1, 1.5, 3, while lines refer to $Z_w =$ 5 (reported as a reference). Data are reported for selected $N_s$, namely $N_s=2,20,32,38,66$, and 90. Panels ($h_1$)-($h_3$) are snapshots of the system with $Z_w =$ 3 at different number of stripes: ($h_1$) $N_s =$ 2, ($h_2$) $N_s =$ 32, ($h_3$) $N_s =$ 90.}
	\label{fig:surfcharge_n2}
\end{figure*}

\begin{figure*}[t]
	\begin{center}
	\includegraphics[width=0.75\textwidth]{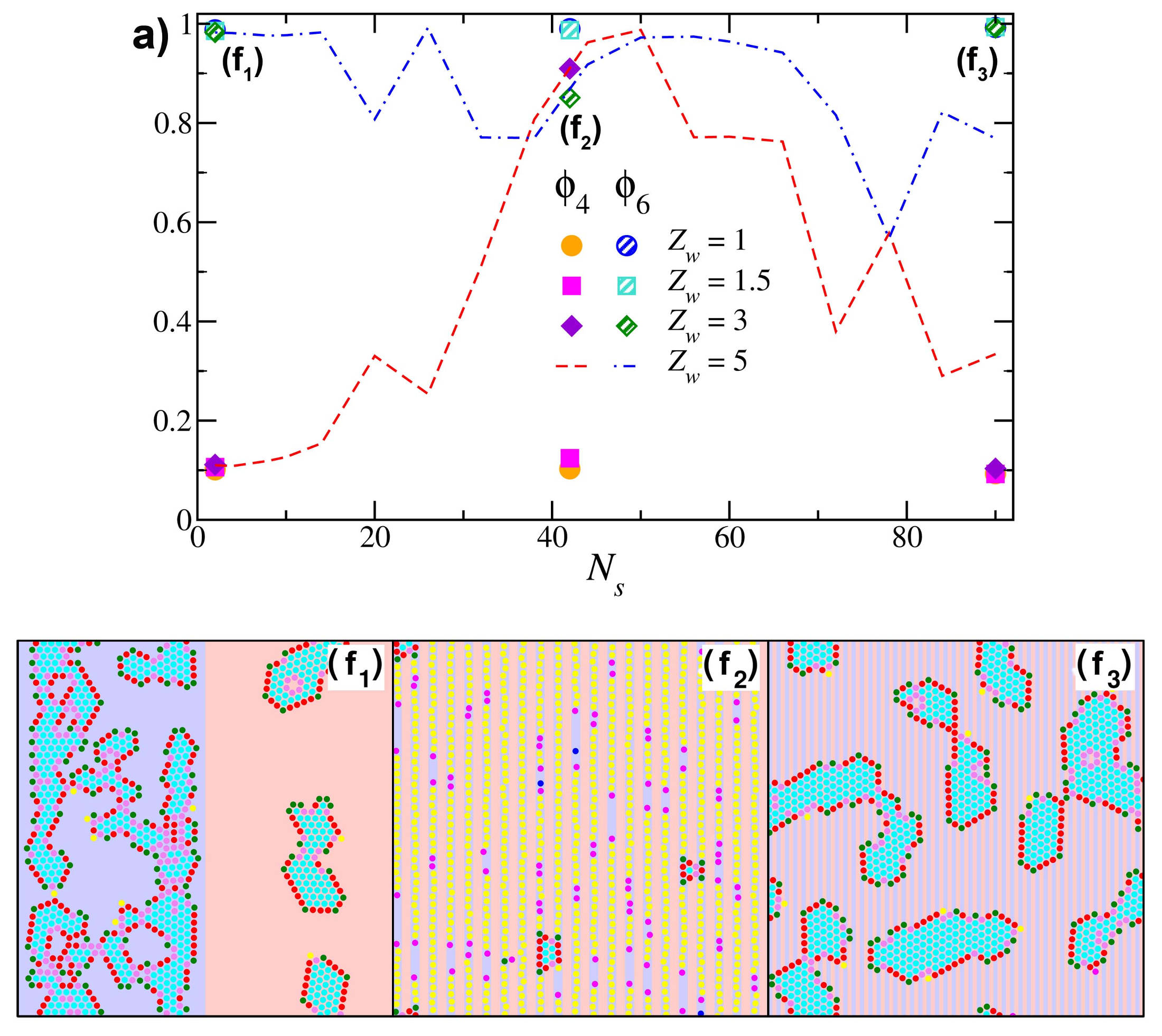} 
	\end{center}
	\caption{System 60c substrate +/-: local bond order parameters, $\phi_4$f and $\phi_6$, as function of $N_s$ for different $Z_w$-values (as labeled): symbols refer to $Z_w =$ 1, 1.5, 3, while lines refer to $Z_w =$ 5 (reported as a reference). Data are reported for selected $N_s$, namely $N_s=2,42$, and 90. Panels ($f_1$)-($f_3$) are snapshots of the system with $Z_w =$ 3 at different number of stripes: ($f_1$) $N_s =$ 2, ($f_2$) $N_s =$ 42, ($f_3$) $N_s =$ 90.}
	\label{fig:surfcharge_c}
\end{figure*}

We first address the effect of a substrate carrying a weaker charge, namely we consider $Z_w = \pm \frac{1}{90} Z_p = \pm 1$, $Z_w = \pm \frac{1.5}{90} Z_p = \pm 1.5$ and $Z_w = \pm \frac{3}{90} Z_p = \pm 3$. Substrates with higher surface charges have not been considered, as both particles 60n and 60c are rather strongly adsorbed on substrates already for $Z_w = \pm 5$. We perform simulations at selected $N_s$-values and selected surface patterns for both systems: namely, we consider system 60n on a +/0 substrate with $N_s=2,38,66$ and 90, system 60n on a +/- substrate with $N_s=2,20,32,38,66$ and 90, and system 60c on a +/- substrate with $N_s=2,42$ and 90. Clearly, at low $Z_w$-values the assembly is mostly driven by the particle-particle interaction, leading to the formation of clusters with triangular arrangements, as this is the preferred configuration in case of a neutral ($i.e.,$ non-adsorbing) substrate for both 60n and 60c systems (see panels a) in Figs.~\ref{fig:surfcharge_n1},~\ref{fig:surfcharge_n2} and~\ref{fig:surfcharge_c}): in all investigated cases, $\phi_4 \approx 0$ and $\phi_6 \approx 1$ over the whole $N_s$-range as long as $Z_w < 3$. As soon as $Z_w \geq $ 3, all systems start to be affected by the substrate, as described in the following. 

We start our discussion with system 60n on a +/0 substrate (see Fig.~\ref{fig:surfcharge_n1}). In this case, the bond order parameters at $N_s=2$ for $Z_w=3$ and $Z_w=5$ are exactly the same, meaning that when particles assemble over extended areas of the substrate no significative effect occurs as long as $ 3 \lesssim Z_w \lesssim 5$. This is also supported by a visual comparison of the snapshots (see panel ($b_1$) of Fig.~\ref{fig:60n_p0} and panel ($g_1$) of Fig.~\ref{fig:surfcharge_n1}). When $N_s$ is increased, though, we observe that a less charged substrate favors the formation of triangular domains: even though $\phi_4$ and $\phi_6$ are less significative in the intermediate $N_s$-range (due to the reduced number of large compact clusters), it is worth noting that $\phi_4 \ll \phi_6$ at $N_s$ = 38 and 66 for $Z_w=3$, while for $Z_w=5$ $\phi_4 \approx \phi_6$ at the same $N_s$-values; the relative abundance of triangular particle arrangements for $Z_w=3$ can be also inferred by a visual inspection of the system (see panel ($b_6$) of Fig.~\ref{fig:60n_p0} and panel ($g_2$) of Fig.~\ref{fig:surfcharge_n1}). This effect can be related to a reduced preference of the particles to sit on the positive stripes, where they arrange in square-like aggregates, thus bringing bonding and adsorption on a comparable energetic ground. The competition between triangular and square particle arrangements is also observed at $N_s=90$, where the formation of the pattern-induced square domains is hindered by the persistence of some triangular domains (see panel ($b_8$) of Fig.~\ref{fig:60n_p0} and panel ($g_3$) of Fig.~\ref{fig:surfcharge_n1}), leading to a decreased (increased) $\phi_4$ ($\phi_6$) for $Z_w=3$. 

If we now consider system 60n on a +/- substrate (see Fig.~\ref{fig:surfcharge_n2}) no significative difference can be found between $Z_w=3$ and $Z_w=5$ as long as $N_s \lesssim 38$. In contrast, at $N_s=90$, we observe a striking difference between the two cases: while for $Z_w=5$ no aggregation occurs, for $Z_w=3$ particles mostly assemble into triangular clusters. In the first case, adsorption to the negative stripes is preferred over bonding: as soon as the interaction with the substrate is weak, the particle-particle based assembly sets in. Here we do not discuss the effect of $Z_w$ on the network forming region at intermediate $N_s$-values as first a deeper analysis would be needed to elucidate the properties of the emerging networks, but this is out of the scope of the present paper.

Finally, we consider system 60c on a +/- substrate (see Fig.~\ref{fig:surfcharge_c}): similar to the 60n system on a +/- substrate, we observe that triangular domains are favored when $Z_w=3$. While for $Z_w=5$ we observe triangular domains at low $N_s$-values (see snapshot ($e_1$) in Fig.~\ref{fig:60c_opposite}) and a coexistence between triangular and square domains at large $N_s$-values (see snapshot ($e_8$) in Fig.~\ref{fig:60c_opposite}), for $Z_w=3$ we observe triangular domains at  both low and large $N_s$-values (see the bond order parameters in panel a) of Fig.~\ref{fig:surfcharge_c} and related snapshots ($f_1$) and ($f_3$)), with no presence of square-like domains at $N_s=90$.

In summary, when the surface charge of the substrate is reduced, bonding is favored with respect to adsorption and thus, the investigated systems tend to assemble into those crystalline domains that emerge spontaneously on a homogeneous, neutral substrate. For systems 60n and 60c on a +/- substrate, the threshold between the two regimes is $Z_w \approx 3$: at such a value of $Z_w$ system 60n on a +/- substrate forms triangular clusters instead of monomers, while system 60c forms triangular domains instead of a mixed phase of triangles and squares. For system 60n on a +/0 substrate in order to observe the triangular domains in place of the pattern-induced square aggregates the surface charge must be very low, $i.e.$, $Z_w \ll 3$.

\subsubsection{Effect of interaction range}

We address here the effect of small variations in the interaction range of system 60n. To this aim, we introduce two new systems: they are both neutral IPCs with patch size defined by $\cos \gamma \approx 60 \degree$ and they differ in their interaction range. The system labelled as SR (short range) has $\delta = $ 0.15, the system labelled LR (long range) has $\delta$ = 0.25. We perform Monte Carlo simulations for  system 60n on both +/0 and +/- substrates at selected state points, namely at $N_s =$ 2, 26, 38, 60, 90.

\begin{figure*}[t]
	\begin{center}
	\includegraphics[width=0.75\textwidth]{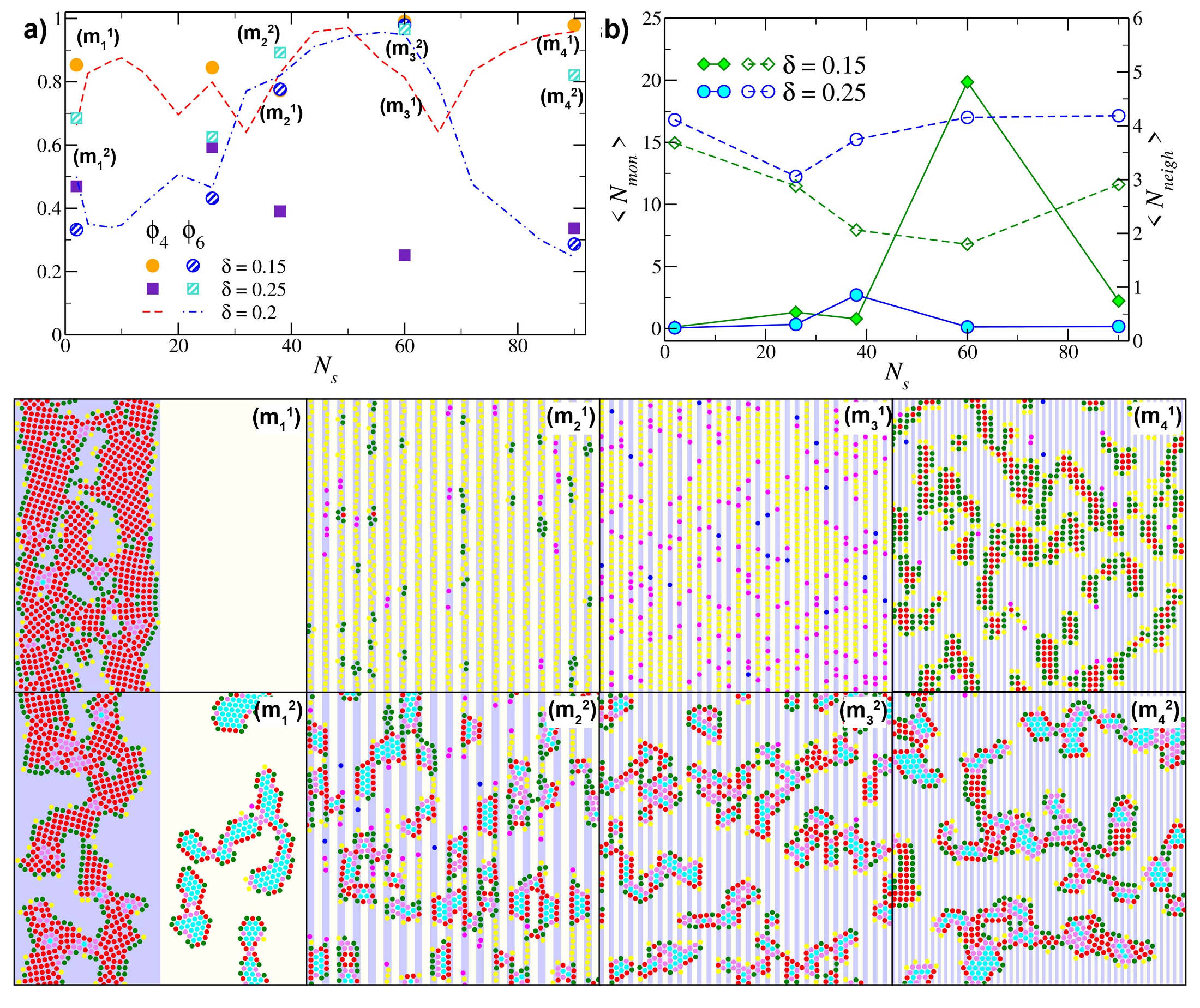} 
	\end{center}
	\caption{a) Bond order parameters $\phi_4$ and $\phi_6$ for systems 60n (reported as a reference), SR and LR (as labelled) on a +/0 substrate as function of the number of stripes $N_s$ . b) Average number of monomers $\langle N_{mon} \rangle$ (full symbols) and average number of neighbors $\langle N_{neigh} \rangle$ (empty symbols) as function of $N_s$ for systems SR and LR (as labelled). Panels ($m^1_1$)-($m^1_4$) are snapshots of SR systems, ($m^2_1$)-($m^2_4$) are snapshots of LR systems, both at different number of stripes: ($m^1_1$) ($m^2_1$) $N_s =$ 2, ($m^1_2$) ($m^2_2$) $N_s =$ 38, ($m^1_3$) ($m^2_3$) $N_s =$ 60, ($m^1_4$) ($m^2_4$) $N_s =$ 90.}
	\label{fig:delta_p0}
\end{figure*}

\begin{figure*}[t]
	\begin{center}
	\includegraphics[width=0.75\textwidth]{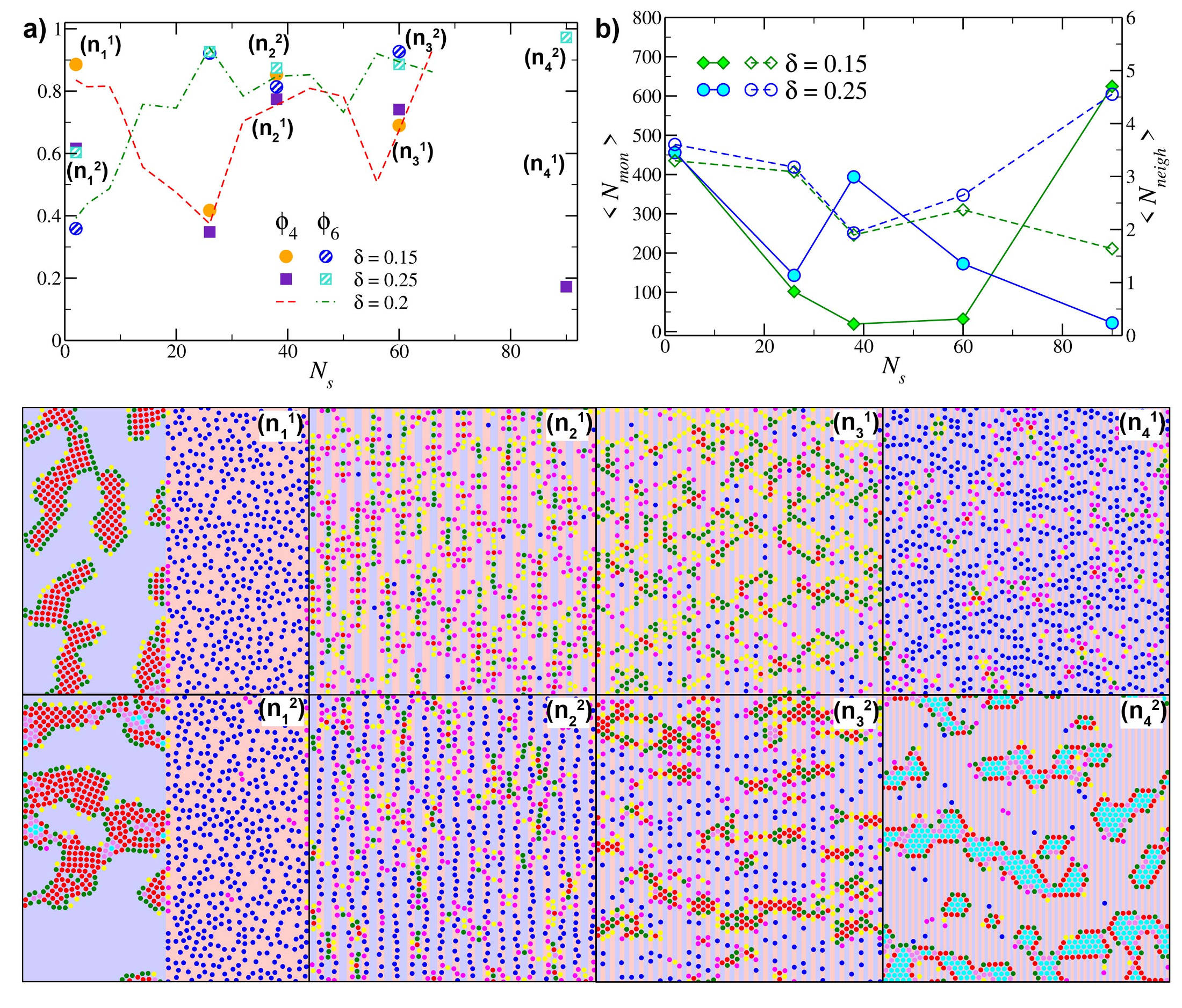} 
	\end{center}
	\caption{a) Bond order parameters $\phi_4$ and $\phi_6$ for systems 60n (reported as a reference), SR and LR (as labelled) on a +/- substrate as function of the number of stripes $N_s$ . b) Average number of monomers $\langle N_{mon} \rangle$ (full symbols) and average number of neighbors $\langle N_{neigh} \rangle$ (empty symbols) as function of $N_s$ for systems SR and LR (as labelled). Panels ($n^1_1$)-($n^1_4$) are snapshots of SR systems, ($n^2_1$)-($n^2_4$) are snapshots of LR systems, both at different number of stripes: ($n^1_1$) ($n^2_1$) $N_s =$ 2, ($n^1_2$) ($n^2_2$) $N_s =$ 38, ($n^1_3$) ($n^2_3$) $N_s =$ 60, ($n^1_4$) ($n^2_4$) $N_s =$ 90.}
	\label{fig:delta_pn}
\end{figure*}

In Figure~\ref{fig:delta_p0}, we report the behavior of both SR and LR systems on a +/0 substrate, together with the reference 60n system. We observe that system SR behaves very similar to system 60n: square domains emerge both at low and high $N_s$-values, $\langle N_{mon} \rangle$ is negligible over the whole $N_s$-range with a mild peak at $N_s \approx 60$ where particles are organized in single-particle strings, while $\langle N_{neigh} \rangle$ fluctuates between four and 3.5 (see panels a) and b) of Fig.~\ref{fig:delta_p0} and the snapshots ($m^1_1$)-($m^1_4$)). In contrast, system LR displays a different assembly scenario. In this case, for $N_s \lesssim 20$, $\phi_4 \approx \phi_6$: a visual inspection of the configurations shows the presence of square domains on the positive stripes and triangular domains on the neutral stripes, whereas at shorter interaction ranges the neutral stripes were completely depleted. This observation already suggests that the adoption to the positive stripes at large $N_s$-values is less effective, thus leading to a competition between (less favored) square-like particles (adsorbed on positive stripes) and triangular-like particles (emerging on neutral -- non adsorbing -- stripes). On further increasing $N_s$ the tendency of the particles to form domains with triangular symmetry persists and increases, as signaled by the increase of $\langle N_{neigh} \rangle$ above four at large $N_s$-values and by the visual analysis (see snapshots($m^2_2$)-($m^2_3$) in Fig.~\ref{fig:delta_pn}). At $N_s =$ 90, we observe a competition between triangular and square-induced arrangements (see snapshot ($m^2_4$) in Fig.~\ref{fig:delta_pn}), while for system 60n and SR the latter
dominates.

In Figure \ref{fig:delta_pn}, we report the behavior of both system SR and LR on a +/- substrate (again the case 60n is reported for comparison). We observe here a scenario similar to the one described for the +/0 case: system SR and 60n behaves in the same way, whereas LR displays a different assembly scenario as soon as $N_s \gtrsim$ 30. In particular, system LR does not form a proper network phase at any $N_s$ investigated, but it tends to form more compact clusters that span over many consecutive stripes, as shown in the snapshot ($n^2_3$) of Fig.~\ref{fig:delta_pn}. Moreover, for very high $N_s$, LR particles assemble into clusters characterized by triangular domains instead of dispersing as monomers on the negative stripes, in contrast to what happens at this $N_s$ for both 60n and SR.\\

In general, longer interaction ranges favor a particle-particle interaction driven assembly and thus the formation of triangular domains, while shorter interaction ranges favor adsorption with respect to bonding.

\subsection{Visual inspection of the configurations}
\label{sec:visual}
So far we have discussed our results in terms of the spacial arrangement of the particles. Given the anisotropy of the particle surface, a closer look at the orientational particle arrangements can provide additional insight into the behavior of the investigated systems. Here we do not perform a quantitative analysis of the particle orientations -- since it is beyond the scope of this paper -- but rather highlight the trends that can be observed in the bonding patterns of the different aggregation regimes described in sections~\ref{sec:distr},~\ref{sec:bondorder} and~\ref{sec:var}. 

As described in section~\ref{sec:model&methods}, we consider IPCs in a quasi two-dimensional slab: particles are thus free to rotate in three dimensions. If we define the particle orientation as the vector joining the particle center and one of the two patches, we have that isolated particles orient themselves (i) parallel to the $xy$-plane on a positive surface or (ii) perpendicular to the $xy$-plane on a negative surface; as soon as the particle interact with other IPCs its orientation is the result of a trade-off between the bonding energy with its neighbors and the bonding energy with its substrate.
A detailed analysis of the energy balance between these two contributions has been carried for systems 60n and 60c on homogeneous substrates in Ref.~\cite{bianchi:2d2014}. Here we briefly summarize the bonding patterns observed in Ref.~\cite{bianchi:2d2014} for the purpose of our discussion (see Fig.~\ref{fig:old} as reference): in the monomer phase particles are perpendicular to the $xy$-plane; in square domains particles bond in a T-shape fashion with polar and equatorial regions of neighboring particles at contact; in triangular domains two types of bonding patterns emerge, defined as flower-like and grain-like patterns. Flower-like patterns emerge in system 60n and are characterized by rings of six particles oriented in the $xy$-plane with one up-right particle in their center. Grain-like patterns characterize system 60c and consist of triangular domains where particles are predominantly oriented along the horizontal direction and form an angle of approximately $60^{\degree}$ with the symmetry axes of the neighboring particles. 

\begin{figure}[htbp]
	\begin{center}
	\includegraphics[width=0.4\textwidth]{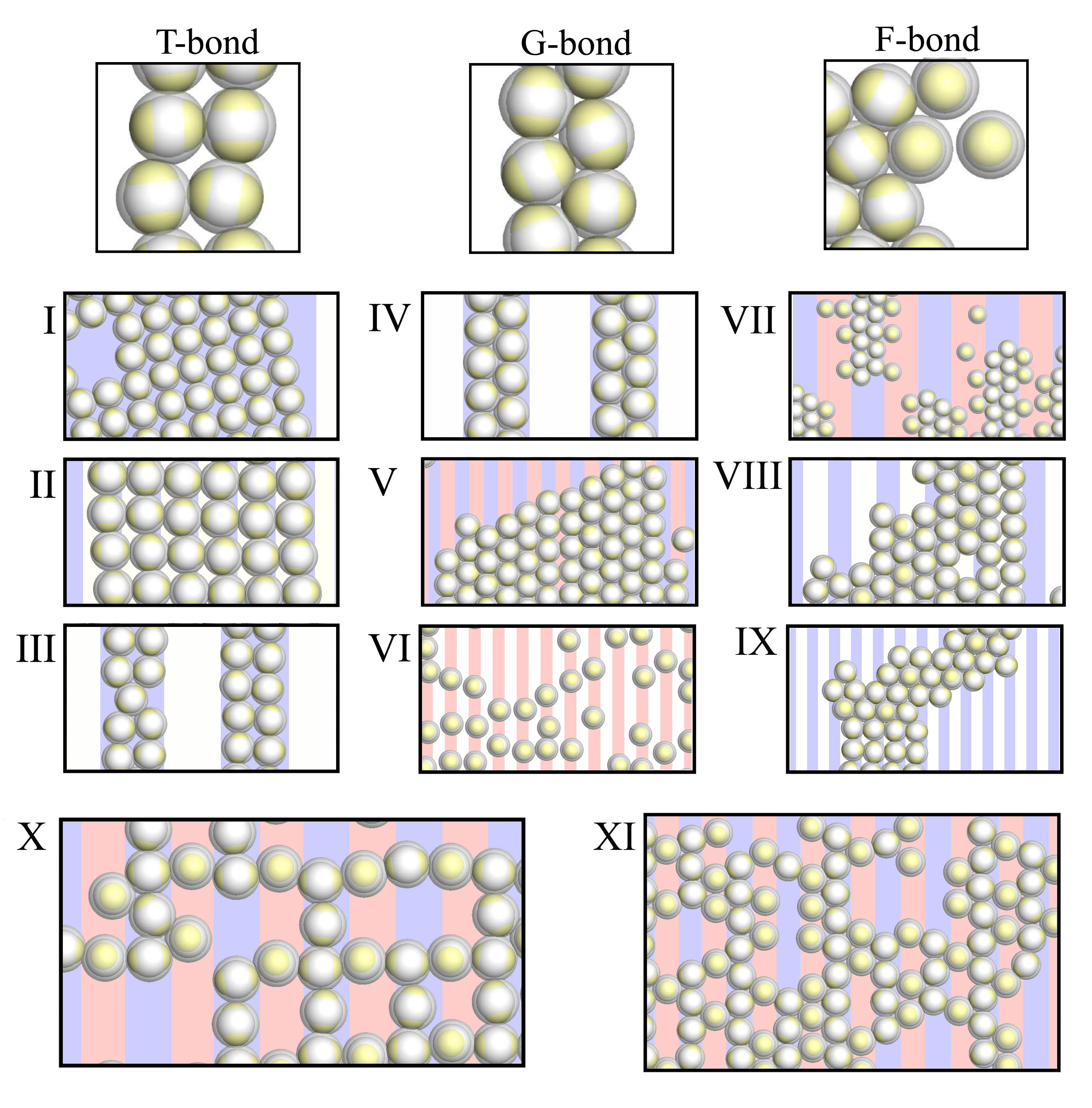}            
		\end{center}
	\caption{Particles in different bonding configurations. Upper row, from left to right: T-like bonds, F-like bonds and G-like bonds. I: Interaction-induced square domains (from snapshot ($\rm b_2$) of Fig.~\ref{fig:60n_p0}). II: Substrate-induced, square domains (from snapshot ($\rm b_8$) of Fig.~\ref{fig:60n_p0}). III: Double-particle lanes with T-like bonds (from snapshot ($\rm b_3$) of Fig.~\ref{fig:60n_opposite}). IV: Double-particle lanes with F-like bonds (from snapshot ($\rm c_3$) of Fig.~\ref{fig:60n_opposite}). IV: Double-particle lanes with G-like bonds (from snapshot ($\rm d_3$) of Fig.~\ref{fig:60n_p0}). V: Mixed square and G-like triangular domains (from snapshot ($\rm e_6$) of Fig.~\ref{fig:60n_opposite}). VI: Monomers (from snapshot ($\rm a_4$) of Fig.~\ref{fig:60n_n0}). VII: F-like triangular domains (from snapshot ($\rm g_2$) of Fig.~\ref{fig:delta_p0}), XI: Mixed square and F-like triangular domains (from snapshot ($\rm m_4^2$ in Fig.~\ref{fig:delta_p0}). X and XI: Networks for system 60n on substrate +/- (from snapshots ($c_5 $) and ($c_6$) of Fig.~\ref{fig:60n_opposite}).
	}
	\label{fig:orientations}
\end{figure}

In view of the previous analysis we define here three types of bonding between IPCs (see top three panels in Fig.~\ref{fig:orientations}): T-bonds, G-bonds and F-bonds. 

T-bonds are in-plane equatorial-polar contacts, they occur when the adsorption on positive stripes drives the assembly of the systems: when particles belong to compact clusters, then the T-like configuration leads to four bonds per particle (see panels I and II of Fig.~\ref{fig:orientations}), while when they belong to double-particle lanes, particles in T-like configurations form on average three bonds (see panel III of Fig.~\ref{fig:orientations}). We observe that, in contrast to square domains formed on homogeneous positive substrates (panel I), pattern-induced square domains have a fixed orientation throughout the sample (panel II).

G-bonds are in-plane grain-like contacts, they occur in system 60c, while they do not emerge for system 60n. In Fig.~\ref{fig:orientations} we report a snapshot of double-particles lanes with a G-bonds pattern, while for extended compact domain we refer to Ref.~\cite{bianchi:2d2014}. We observe that, when triangular and square domains coexist for system 60c at large $N_s$-values, G-bonds and T-bonds patterns emerge in the compact clusters, where T-like lanes are adsorbed only on the positive stripes, while G-like lanes can be on both stripe-types (see panel V of Fig.~\ref{fig:orientations}).

Finally, F-bonds are out-of-plane equatorial-polar -- or flower-like -- contacts, they occur in system 60n and they are responsible for a large range of aggregation scenarios, as discussed in the following.  This is due to the fact that 60n particles on a negative substrate tend to orient themselves up-right (see panel VI of Fig.~\ref{fig:orientations}), possibly acting as bridges between particles adsorbed in-plane. In particular, for system 60n on a +/- substrate, when the stripes can accomodate exactly two particles, the T-bonds pattern on the positive stripes disappears in favor of double-particle lanes in a G-like fashion, where F-like particles adsorbed on the negative stripes set the boundaries of the clusters: two bridge-particles in an up-right position can sit on the same stripe, thus effectively forcing the clusters to grow along the $y$ direction (see panel VIII of Fig.~\ref{fig:orientations}). When stripes are such that a single particle cannot be fully embedded, we observe either F-like extended domains (on a +/- substrate) or a coexistence between square and F-like triangular domains (on a +/0 substrate). In both cases up-right particles lie on the negative substrate (see panel IX of Fig.~\ref{fig:orientations}) and act as bridge between two G-like compact clusters, thus effectively allowing the growth of the clusters along the $x$ direction. 
Finally, the interplay between the different anisotropic interactions and the length scales in system 60n can give rise to even more complex scenarios: panels X and XI of Fig.~\ref{fig:orientations} show examples of open clusters, characterized by strings of particles with alternating up-right and planar orientation. The structure of these networks seems to be very sensitive to the substrate pattern and a more focused investigation is needed. 

\section{Conclusions}\label{sec:conclusions}

Within the framework of materials design, we have investigated how to create colloidal monolayers on substrates by tuning the surface properties of both the particles and the substrate. In particular, we have considered colloids with a charged, non-homogeneous surface close to a planar wall characterized by parallel stripes with alternating surface charge. By tuning the competition between the particle-particle and the particle-substrate interaction as well as between the particle size and the width of the stripes, we have shown how to control the spacial ordering of the colloids and how to drive the systems from compact, to elongated or open aggregates. 

We have considered negatively charged colloids with two positively charged polar regions, the size of these regions being relatively large, the net charge of the particles being zero or slightly negative, with a relatively short interaction range. These systems were shown to give rise to a rich assembly scenario already close to a homogeneous substrate: crystalline domains with different spatial (square $versus$ triangular) arrangements or different orientational bonding (flower-like $versus$ grain-like) patterns as well as monomers with a well-defined orientation were observed on changing the charge ratios between the differently charged surfaces. Here we have investigated the effect of properly designed substrate motifs. In particular, we have considered neutral, positive or negative stripes in an alternating pattern: negative/neutral, positive/neutral or positive/negative.  

One striking effect of the substrate pattern is the robust formation of pattern-induced crystalline domains with square symmetry: these particle arrangements emerge for both overall neutral and charged particles  as soon as (i) particles are preferentially adsorbed on the positive stripes and (ii) the width of the stripes becomes much smaller than the particle size. In some cases, these pattern-induced square domains compete with (flower- as well as grain-like) triangular domains but it is possibile to control the energy balance between adsorption (that favor square domains) and bonding (that favors triangular domains) by, $e.g.$, changing the substrate surface charge or the screening conditions. On reducing the substrate charge, both grain- and flower-like triangular domains are highly favored with respect to square domains (for system 60c and 60n on a +/0) or to monomers (for system 60n on a +/- substrate). Changing the electrostatic screening conditions implies variation in the interaction range: we observe that long range interactions favor particle-particle bonding, while short range interactions enhance particle adsorption. On reducing the (particle-particle and particle-substrate) interaction range, we thus observe that pattern-induced scenarios prevail, meaning that a monomer phase (if particles prefer to adsorb on negative stripes) or a square pattern (if particles tend to adsorb on positive stripes) can prevail over the triangular arrangement, while for longer interaction ranges triangular domains or clusters with mixed symmetry are observed. 

In summary, double particle lanes with square or triangular symmetry, elongated (along the $y$ axis) versus extended (along the $x$ axis) clusters and pattern-induced crystals with a well-defined orientation can be induced by playing with the competition between  different anisotropic interactions and length scales. The balance between adsorption and bonding is affected my the many parameters of the systems and the extended simulations carried here show that non-intuitive behaviors can emerge.  

We have focused mostly on the spacial arrangement of the particles. Future work should address, $e.g.,$ the orientational features of the bonding patterns within the clusters, the compactness and the shape of the aggregates, their preferred direction with respect to the substrate motif as well as their percolating properties.

Moreover, it would be interesting to gradually release the confinement towards bulk systems: the aggregation of colloidal particles on patterned substrates could be used to guide the crystallization of bulk colloidal crystals as well as to tailor the orientation and size of the resulting lattices. This process is referred to as ``colloidal epitaxy"~\cite{colloidalepitaxy,Dias2013,Dias2017}. We note, for instance that, in contrast to the square arrangements assembled on a homogeneous positive substrate, pattern-induced square domains have a well-defined crystallographic axis, which may be used as templates to assemble perfect square crystals on large (possibly macroscopic) scales.

\section{Acknowledgements}
The authors wishes to thank Benedikt Vitecek for preliminary results. EB acknowledges support from the Austrian Science Fund (FWF) under Proj. Nos. V249-N27. Computer time at the Vienna Scientific Cluster (VSC) is also gratefully acknowledged.

\bibliographystyle{rsc}
\bibliography{ipc-patterns-biblio}

\end{document}